# Astronomical Detection of the Interstellar Anion C₁₀H⁻ towards TMC-1 from the GOTHAM Large Program on the GBT

Anthony Remijan ,[1] Haley N. Scolati ,[2] Andrew M. Burkhardt ,[3] P. Bryan Changala,[4]
Steven B. Charnley,[5] Ilsa R. Cooke,[6] Martin A. Cordiner,[5, 7] Harshal Gupta ,[4, 8] Eric Herbst,[2, 9]
Kin Long Kelvin Lee ,[10, 11, 4] Ryan Loomis,[1] Christopher N. Shingledecker ,[12] Mark A. Siebert,[9] Ci Xue,[11]
Michael C. McCarthy,[4] and Brett A. McGuire [11, 1]

[1]*National Radio Astronomy Observatory, Charlottesville, VA 22903, USA*
[2]*Department of Chemistry, University of Virginia, Charlottesville, VA 22904, USA*
[3]*Department of Physics, Wellesley College, Wellesley, MA 02481, USA*
[4]*Center for Astrophysics | Harvard & Smithsonian, Cambridge, MA 02138, USA*
[5]*Astrochemistry Laboratory and the Goddard Center for Astrobiology, NASA Goddard Space Flight Center, Greenbelt, MD 20771, USA.*
[6]*Department of Chemistry, University of British Columbia, 2036 Main Mall, Vancouver BC V6T 1Z1, Canada*
[7]*Institute for Astrophysics and Computational Sciences, The Catholic University of America, Washington, DC 20064, USA*
[8]*National Science Foundation, Alexandria, VA 22314, USA*
[9]*Department of Astronomy, University of Virginia, Charlottesville, VA 22904, USA*
[10]*Accelerated Computing Systems and Graphics Group, Intel Corporation, 2111 NE 25th Ave, Hillsboro, OR 97124, USA*
[11]*Department of Chemistry, Massachusetts Institute of Technology, Cambridge, MA 02139, USA*
[12]*Department of Physics and Astronomy, Benedictine College, Atchison, KS 66002 USA*

## ABSTRACT

Using data from the GOTHAM (GBT Observations of TMC-1: Hunting for Aromatic Molecules) survey, we report the first astronomical detection of the C₁₀H⁻ anion. The astronomical observations also provided the necessary data to refine the spectroscopic parameters of C₁₀H⁻. From the velocity stacked data and the matched filter response, C₁₀H⁻ is detected at $>9\sigma$ confidence level at a column density of $4.04^{+0.67}_{-2.23} \times 10^{11}$ cm⁻². A dedicated search for the C₁₀H radical was also conducted towards TMC-1. In this case, the stacked molecular emission of C₁₀H was detected at a $\sim3.2\sigma$ confidence interval at a column density of $2.02^{+2.68}_{-0.82} \times 10^{11}$ cm⁻². However, since the determined confidence level is currently $<5\sigma$, we consider the identification of C₁₀H as tentative. The full GOTHAM dataset was also used to better characterize the physical parameters including column density, excitation temperature, linewidth, and source size for the C₄H, C₆H and C₈H radicals and their respective anions, and the measured column densities were compared to the predictions from a gas/grain chemical formation model and from a machine learning analysis. Given the measured values, the C₁₀H⁻/C₁₀H column density ratio is $\sim2.0^{+5.9}_{-1.2}$, the highest value measured between an anion and neutral species to date. Such a high ratio is at odds with current theories for interstellar anion chemistry. For the radical species, both models can reproduce the measured abundances found from the survey; however, the machine learning analysis matches the detected anion abundances much better than the gas/grain chemical model, suggesting that the current understanding of the formation chemistry of molecular anions is still highly uncertain.

*Keywords:* Astrochemistry, ISM: molecules

## 1. INTRODUCTION

It has now been more than 15 years since the first reported molecular anion was detected in astronomical environments with the discovery of C₆H⁻ (McCarthy et al. 2006). This detection was the culmination of the early predictions that anions should be present under interstellar conditions (Herbst 1981). And, as summarized by Millar et al. (2017) molecular anions are crit-ically important to the physics and evolution of astronomical objects - including building up structures in the early universe, dominating the visible opacity for stars like the sun, being the possible carriers to the diffuse interstellar bands and in determining the physical and chemical environments of astrophysical regions including measuring the impact of interstellar radiation fields and other molecular cloud properties given their reactivity. As such, the detection of interstellar molecular anions are far beyond just an astrochemical curiosity.

The initial detection then sparked the search for new molecular anions and the identification in $C_4H^-$ (Cernicharo et al. 2007; Agúndez et al. 2008), $C_8H^-$ (Brünken et al. 2007; Remijan et al. 2007; Gupta et al. 2007; Kawaguchi et al. 2007), $C_3N^-$ (Thaddeus et al. 2008), $C_5N^-$ (Cernicharo et al. 2008, 2020) and $CN^-$ (Agúndez et al. 2010), observed primarily toward the dark cloud source TMC-1 and the evolved star IRC+10216. Since these initial detections, searches for $C_6H^-$ have shown that this anion is abundant in a variety of sources, from quiescent dark clouds to active star-forming regions (Cordiner et al. 2013). These searches revealed a wide discrepancy in the anion-to-neutral column density ratios. The anion-to-neutral column density ratio for carbon chains varies markedly with chain length and the astrophysical environment. In TMC-1, the $C_6H^-$ /$C_6H$ ratio is $\sim$2.5%, whereas the $C_4H^-$ /$C_4H$ ratio is only $\sim$0.0012% (Cordiner et al. 2013), increasing 20-fold to $\sim$0.024% towards IRC+10216 (Cernicharo et al. 2007). The enhanced anion fraction of longer chains is mirrored by the N-terminated species: $C_5N^-$ /$C_5N$ $\approx$ 12.5% and $C_3N^-$ /$C_3N$ $\approx$ 0.7% in TMC-1. (Cernicharo et al. 2020). These observations highlight that anion chemistry and the limit to which carbon chain anions can grow in astronomical environments are still not well understood.

Recently, Siebert et al. (2022) reported the identification of the largest $CH_3$-terminated carbon chain molecule $CH_3C_7N$ toward TMC-1 and in that work, determined the column densities of several large carbon chain families and made predictions for the column densities for those species yet to be detected. These predictions were made using the three-phase gas–grain chemical network model nautilus (v1.1, Ruaud et al. 2016). In addition, a linear extrapolation as a function of chain length was made from the measured column densities of the smaller carbon chains. While the predicted and measured column density ratios can differ up to an order of magnitude, it served as motivation to conduct an astronomical search for the elusive, larger carbon chain species that may already be contained within our existing TMC-1 dataset.

As such, to follow on from the recent detections and surveys of molecular anions, an extensive search for the decapentaynyl radical ($C_{10}H$) and the decapentaynyl anion ($C_{10}H^-$ ) was conducted toward the dark cloud TMC-1 — the site of the first detection of interstellar anions — with the GOTHAM (GBT Observations of TMC-1: Hunting for Aromatic Molecules) survey. The family of H-terminated carbon chain radicals - $C_{2n}H$ (where n > 1) and their associated anions have been studied in various astronomical environments and these species are believed to be share common chemical formation pathways (Bettens & Herbst 1997; Walsh et al. 2009). To best constrain the column density ratios between the radical and anion species, we performed a complete re-analysis of $C_4H$, $C_6H$, $C_8H$ radicals and their associated anions taking advantage of the significantly improved signal-to-noise ratio and spectral resolution of our data compared to their original detections. These observations provide the most rigorous measurement of the column densities of these species with which to compare to chemical formation and machine learning models. They have also set a new limit to the largest carbon chain species detectable in astronomical environments as the first detection of $C_{10}H^-$ and a tentative detection of $C_{10}H$ are reported towards TMC-1. The observing parameters describing the search for these carbon-chain species are presented in Section 2. The new spectroscopic analyses for $C_4H$, $C_6H$, and $C_{10}H^-$ are given in Section 3, which includes a discussion of how the astronomical detection of $C_{10}H^-$ enabled the more accurate determination of its molecular constants. The observational analyses on how the physical parameters of TMC-1 are determined from these data are presented in Section 4. The results of the observational searches for $C_{10}H$ and $C_{10}H^-$ and the previously detected carbon-chain molecules are presented in Section 5. A comparison of the predicted-to-measured column density ratios for this family of carbon-chain molecules from both state-of-art gas/grain chemical models and from machine learning analyses are given in Section 6. Finally, our conclusions are highlighted in Section 7.

## 2. OBSERVATIONS

Observations for this study were obtained as part of the GOTHAM Survey. GOTHAM is a large project on the 100m Robert C. Byrd Green Bank Telescope (GBT) currently managed by the Green Bank Observatory (GBO). The GOTHAM program is a dedicated spectral line observing program of TMC-1 covering almost 30 GHz of bandwidth at high sensitivity and spectral resolution. All data were taken with a uniform frequency resolution of 1.4 kHz (0.05–0.01 km/s in veloc-ity) and an RMS noise of $\sim$2–20 mK across most of the observed frequency range with the RMS gradually increasing toward higher frequency because of the shorter integration times. This work uses the fourth data reduction (DR4) of GOTHAM targeting the cyanopolyyne peak (CP) of TMC-1, centered at $\alpha_{J2000} = 04^h 41^m 42.5^s$, $\delta_{J2000} = +25°41'26.8''$. Briefly, the spectra in these data cover the entirety of the X-, K-, and Ka-receiver bands with nearly continuous coverage from 7.9 to 11.6 GHz, 12.7 to 15.6 GHZ, and 18.0 to 36.4 GHz (24.9 GHz of total bandwidth). Data reduction involved removal of RFI and artifacts, baseline continuum fitting, and flux calibration using complementary VLA observations of the source J0530+1331. Uncertainty from this flux calibration is estimated at $\sim$20%, and is factored into our statistical analysis described below (McGuire et al. 2020a). A full description of the fourth data reduction can be found in Sita et al. (2022) and the observing strategy and reduction pipeline is fully described in McGuire et al. (2020a).

## 3. SPECTROSCOPIC ANALYSES

Upon comparison of the astronomical data and the fitted laboratory spectra found in the publically available Cologne Database for Molecular Spectroscopy (CDMS; Endres et al. 2016), it was determined that a re-analysis of the measured and predicted frequencies was needed for $C_4H$ and $C_6H$. The spectroscopic data available for the other radicals and associated anions very closely matched the observational spectrum and no re-analysis was required. The subsections below describe the process for how the catalogs for $C_4H$ and $C_6H$ were revised and how the catalog for $C_{10}H^-$ was generated using the astronomically measured spectroscopic data.

### 3.1. $C_4H$

Upon inspection of the signal from $C_4H$ in the GOTHAM data, it was immediately obvious that the available line frequencies in the Cologne Database for Molecular Spectroscopy (CDMS; Endres et al. 2016) were insufficiently accurate to reproduce the observations. In particular, the hyperfine splitting, which is well-resolved in the GOTHAM data, was poorly matched by the catalog data. In some cases, the predicted line frequencies varied by more than a full-width half-maximum linewidth compared to the astronomically detected features, thus necessitating new high-resolution measurements.

$C_4H$ was prepared in the laboratory using the same methods as previous experiments on its $^{13}C$ isotopologues (Chen et al. 1995) and vibrationally excited states (Cooksy et al. 2015). A mixture of 0.1% acetylene was seeded in neon at a pressure of 2.5 kTorr and supersonically expanded along the axis of a cavity Fourier transform microwave (FTMW) spectrometer (Grabow et al. 2005) in 400 $\mu$s gas pulses at a rate of 6 Hz. During each gas pulse, a 1.0 kV discharge was struck between two copper electrodes placed immediately after the valve aperture, creating reactive products which combined to make, among other species, $C_4H$. Three perpendicular pairs of Helmholtz coils are positioned around the spectrometer and tuned to null Earth's magnetic field to less than 50 milligauss throughout the cavity volume.

A total of 22 hyperfine-resolved transitions between 9 and 38 GHz were measured and assigned. These were combined with previous sub-millimeter measurements of a further 10 spin-rotation transitions between 143 and 200 GHz reported by Gottlieb et al. (1983), for which hyperfine structure was not resolved. The dipole moment, which has been a matter of some debate over the years due to the complicated electronic structure of $C_4H$, was taken to be 2.1 Debye (D), as recently determined using high-level quantum chemical calculations (Oyama et al. 2020). Fitting was performed using the SPFIT/SPCAT suite of programs Pickett et al. (1998). The full measured line list, including which lines are used from which data sources, along with the corresponding input and output files from SPFIT/SPCAT, are provided as Supplemental Information.

### 3.2. $C_6H$

Although not quite as striking as for $C_4H$, it was evident from our high-resolution GOTHAM observations that existing catalogs for $C_6H$ were insufficiently accurate to reproduce the observational data. To address this, we have refit all of the high-resolution experimental lines from the work of Gottlieb et al. (2010), which had not previously been included. The updated catalog was sufficiently accurate for our purposes. The full measured line list, along with the corresponding input and output files from SPFIT/SPCAT, are provided as Supplemental Information.

### 3.3. $C_{10}H^-$

To our knowledge, no laboratory spectra exist for $C_{10}H^-$, and our own efforts to produce detectable quantities in our instruments have not yet been successful. Unlike $C_4H$ and $C_6H$, however, $C_{10}H^-$ is a closed-shell linear molecule, and as such, it is straightforward to predict its rotational spectrum, which presents as a series of lines spaced by $\sim$2$B$. Given a reliable prediction of the rotational constant, $B$, prior work has shown that it is possible to identify such species in interstellar spectra preceding their laboratory confirmation. Ex-amples include $C_3H^+$ (Pety et al. 2012) in the Horsehead PDR and $C_5H^+$ (Cernicharo et al. 2022), $HC_5NH^+$ (Marcelino et al. 2020), and $HC_7NH^+$ (Cabezas et al. 2022) in TMC-1, among others.

We began our search for $C_{10}H^-$ by using a value of $B$ = 299.882 MHz and $D$ = 1 Hz, obtained by extrapolating and scaling from the shorter members of the family of anions. The $B$ value specifically is in excellent agreement with that obtained from a quantum chemical calculation carried out at the M06-2X/6-31+G(d) level of theory and basis set using the Psi4 suite of programs (Smith et al. 2020) of $B$ = 303.761 MHz. This level of theory and basis set has been previously shown to reliably produce rotational constants in excellent agreement with experiment (Lee & McCarthy 2020).

Using our estimated values of $B$ and $D$, we produced a catalog of lines and performed a simulation, using a set of fiducial values for $v_{lsr}$, column density ratios between velocity components, excitation temperature ($T_{ex}$), and linewidth ($\Delta V$). These values are based on our prior detection and treatment of benzonitrile within the source, and have been shown to be excellent starting points for the analysis of other molecules in the GOTHAM data (see Supplementary Information of McGuire et al. 2021). No individual transitions were seen above the noise level of the observations. A spectral stack of the data using this first pass estimated catalog, however, revealed a strong (>5$\sigma$) signal in the stacked spectra shifted by just over 10 km s$^{-1}$ from the expected central velocity (Fig. 1).

We then performed a least-squares fit to determine what value of the rotational constant would be required to reproduce the observed signal at the expected central velocity. We derive a value of $B$ = 299.87133 MHz, differing from the scaled prediction by 11 kHz or 0.004%. This derived value was then used to generate a final spectral line catalog for $C_{10}H^-$ that was used for the remainder of the analysis described in this paper; a summary of the values of the $B$ rotational constant determined by our methods are summarized in Table 1. As described in detail in McGuire et al. (2021), a fractional accuracy of better than 10$^{-5}$ (and ideally better than 10$^{-6}$ in the rotational constants of a molecule is required to recover any significant signal from our spectral stacking techniques. Thus, we can infer that our derived value of $B$ is likely accurate to at least 3 kHz.

## 4. OBSERVATIONAL ANALYSIS

In order to derive physical parameters (column density [$N_T$], excitation temperature [$T_{ex}$], linewidth [$\Delta V$], and source size [″]) for the target molecules in our observations, we used the same Markov Chain Monte Carlo (MCMC) model employed in prior GOTHAM analyses (see, e.g., Sita et al. 2022; Siebert et al. 2022; Lee et al. 2021c) and discussed in detail in Loomis et al. (2021). In short, the MCMC model calculates the probability distributions and co-variances for these parameters which are used to describe the emission of molecules observed in our data. The resulting corner plots for the molecules analyzed here are shown in Figs. A2, A3, B1, C1, D1, E1, F1, and G1.

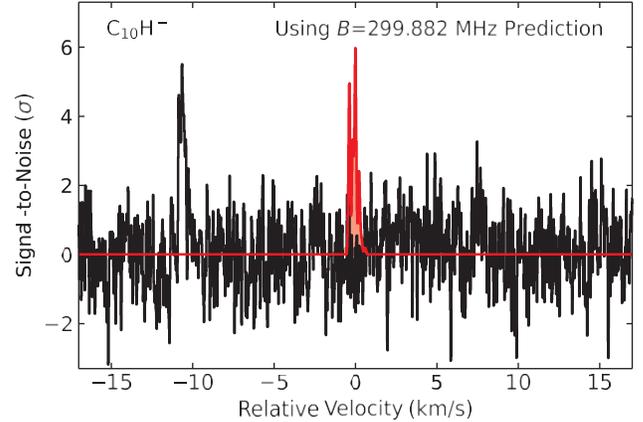

**Figure 1.** Velocity stacked spectra of $C_{10}H^-$ using our first-pass estimated catalog generated with $B$ = 299.882 MHz. The stacked spectra from the GOTHAM DR4 data are displayed in black, overlaid with the expected line profile in red from our first-pass catalog and using our fiducial molecular parameters. The intensity of the simulation is arbitrarily scaled (no fit has been performed). The signal-to-noise ratio is on a per-channel basis.

**Table 1.** Rotational constant of $C_{10}H^-$ from quantum chemistry, scaled from experimental values of smaller anions, and determined in this work from our astronomical observations.

| Parameter | M06-2X/6-31+G(d) | Scaled | This Work |
|---|---|---|---|
| $B$ (MHz) | 301.353 | 299.882 | 299.87133 |
| $D$ (Hz) | - | [1.0]$^†$ | [1.0]$^†$ |

$^†$ $D$ was fixed to a value near that of $C_8H^-$ ($D$ = 4.3 Hz; Gupta et al. 2007), but no attempt was made to refine it further, given the limitations of the analysis.

We adopt the 50$^{th}$ percentile value of the posterior probability distributions as the representative value of each parameter for the molecule and use the 16$^{th}$ and 84$^{th}$ percentile values for the uncertainties. For probabilities that show a Gaussian distribution, these correspond to the 1$\sigma$ uncertainty level. Many of our resulting probability distributions are indeed either Gaussian or nearly Gaussian, and thus these values are usually quite representative of the $1\sigma$ uncertainties. One of the advantages of the MCMC technique over a traditional least-squares fit approach is that far more of parameter space is explored. Correspondingly, a much larger exploration of the uncertainty space is performed as well, including highlighting parameters that may be highly covariant with one another. This manifests as non-separable distributions in the corner plots.

To explore this parameter space with our MCMC approach, a model of the molecular emission is generated for each set of parameters using the molsim software package (Lee et al. 2021a) and following the conventions of Turner (1991) for a single excitation temperature and accounting for the effect of optical depth. Prior observations from GOTHAM (Xue et al. 2020) and others (Dobashi et al. 2018, 2019) have found that most emission seen at centimeter wavelengths in TMC-1 can be separated into contributions from four distinct velocity components within the larger structure, at approximately 5.4, 5.6, 5.8, and 6.0 km s$^{-1}$ (Loomis et al. 2021). In some cases, especially for less abundant species where there is not a clear detection in one of the velocity components, we find that a three-component model has performed better (McGuire et al. 2020b).

To determine the statistical evidence that our model of the emission of these molecules is consistent with the data, we followed the procedures described in detail in Loomis et al. (2021) and performed a spectral stack and matched filtering analysis for $C_{10}H$ and $C_{10}H^-$. Briefly, a weighted average of the observational spectra in velocity space and centered on each spectral line of a target molecule was performed. The weights were determined by the relative intensity of the expected emission (based on the MCMC-derived parameters) and the local RMS noise of the observations. Considering the weak expected intensities for both $C_{10}H^-$ and the $C_{10}H$ isomers, any observational windows containing emission at $>5\sigma$ were ignored in the analysis of those molecules.

Simulated spectra of the molecular emission using the same MCMC-derived parameters were then also generated and stacked using identical weights. This simulation was then used as a matched filter, which is passed through the observational signal. The resulting impulse response function represents the statistical evidence that our model of the emission from the molecule – and thus our derived parameters for the molecule – is consistent with the observations. In addition to the details of the methodology provided in Loomis et al. (2021), the appendices of McGuire et al. (2021) include an extensive analysis of the robustness of the methodology, including the improbability of spurious signals and the minimal impact of red-noise on the procedure.

We also performed MCMC fits for the more abundant $C_4H$, $C_4H^-$, $C_6H$, $C_6H^-$, $C_8H$, and $C_8H^-$ molecules. Given the low line density in our spectra, it is extremely improbable that interfering signals from other species would be present. Still, the spectral regions containing these transitions were manually inspected to ensure there are no interloping signals or other concerns. All strongly detected lines in our data are shown in the Appendix for $C_4H$ (Figs. B2–B4), $C_4H^-$ (Fig. C2), $C_6H$ (Figs. D2 and D3), $C_6H^-$ (Fig. E2), $C_8H$ (Fig. F2), and $C_8H^-$ (Fig. G2).

## 5. OBSERVATIONAL RESULTS

Figure 2 shows the stacked data, the stacked MCMC model, as well as the matched filter response and the first detection of interstellar $C_{10}H^-$ toward TMC-1. The stacked emission in the right panel of Figure 2 exhibits evidence at $>9\sigma$ for the presence of this molecule. Figure A1 shows the individual spectral lines present in the survey. While no individual lines of $C_{10}H^-$ are present above the current noise level of the survey at $>3\sigma$, there are spectral features seen in the corresponding passbands that show some emission above the noise (see, e.g. transitions at 10195.41, 13194.04, 14993.2 and 19191.26 MHz). Table 2 lists the measured physical values determined from the $C_{10}H^-$ fits. In this case, 3 independent velocity components were detected and a total $C_{10}H^-$ column density of $4.04^{+10.67}_{-2.23} \times 10^{11}$ cm$^{-2}$ was determined.

**Table 2.** Summary Statistics of the Marginalized $C_{10}H^-$ Posterior

| $v_{lsr}$ (km s$^{-1}$) | Size ($''$) | $N_T$ (10$^{11}$cm$^{-2}$) | $T_{ex}$ (K) | $\Delta V$ (km s$^{-1}$) |
|---|---|---|---|---|
| $5.624^{+0.012}_{-0.014}$ | $16^{+9}_{-7}$ | $3.68^{+10.67}_{-2.23}$ | | |
| $5.759^{+0.020}_{-0.020}$ | $27^{+10}_{-9}$ | $0.27^{+0.27}_{-0.12}$ | $4.00^{+0.21}_{-0.21}$ | $0.360^{+0.027}_{-0.025}$ |
| – | – | – | | |
| $6.040^{+0.019}_{-0.019}$ | $659^{+234}_{-274}$ | $0.09^{+0.02}_{-0.02}$ | | |
| $N_T$ (Total): $4.04^{+10.67}_{-2.23} \times 10^{11}$ cm$^{-2}$ | | | | |

Figure 3 show the stacked observational data, the stacked MCMC model, as well as the matched filter response and a tentative detection of the $C_{10}H$ radical towards TMC-1. The stacked molecular emission of $C_{10}H$ in the right panel of Figure 3 exhibits evidence at $\sim 3.2\sigma$ for the presence of this molecule; no individual spectral lines were present. Table 3 lists the measured physical values determined from the $C_{10}H$ fits. In this case, 4 independent velocity components were fit and a to-tal $C_{10}H$ column density of $2.02^{+2.68}_{-0.82} \times 10^{11}$ cm$^{-2}$ was reported.

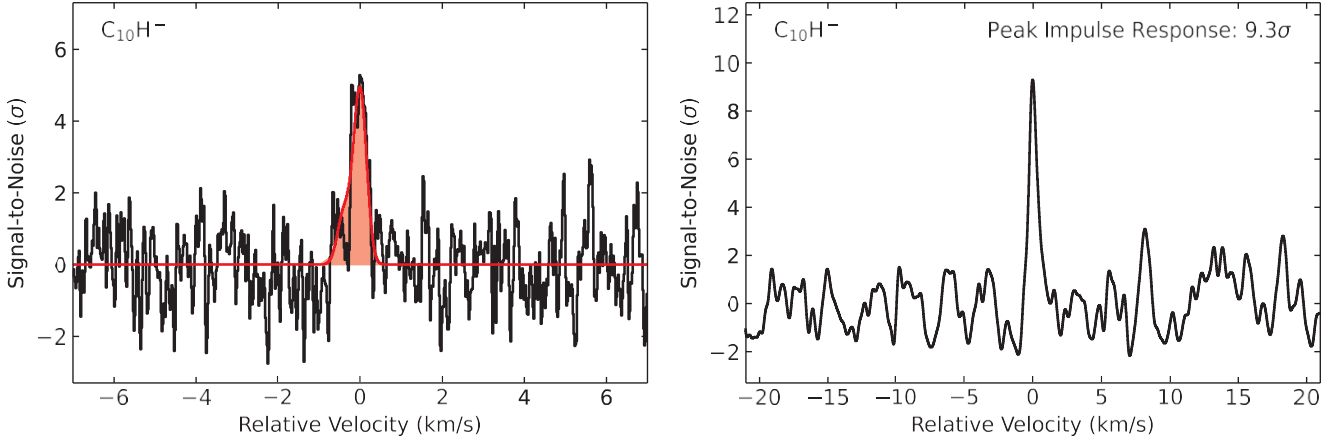

**Figure 2.** Velocity stacked and matched filter spectra of $C_{10}H^-$. The intensity scales are the signal-to-noise ratios (SNR) of the response functions when centered at a given velocity. The "zero" velocity corresponds to the channel with the highest intensity to account for blended spectroscopic transitions and variations in velocity component source sizes. (*Left*) The stacked spectra from the GOTHAM DR4 data are displayed in black, overlaid with the expected line profile in red from our MCMC fit to the data. The signal-to-noise ratio is on a per-channel basis. (*Right*) Matched filter response obtained from cross-correlating the simulated and observed velocity stacks in the left panel; value annotated corresponds to the peak impulse response of the matched filter.

**Table 3.** Summary Statistics of the Marginalized $C_{10}H$ Posterior

| $\upsilon_{lsr}$ (km s$^{-1}$) | Size ($''$) | $N_T$ ($10^{11}$cm$^{-2}$) | $T_{ex}$ (K) | $\Delta V$ (km s$^{-1}$) |
|---|---|---|---|---|
| $5.590^{+0.010}_{-0.009}$ | $24^{+4}_{-4}$ | $0.53^{+0.26}_{-0.16}$ | | |
| $5.726^{+0.015}_{-0.010}$ | $29^{+4}_{-4}$ | $0.29^{+0.13}_{-0.09}$ | | |
| $5.859^{+0.010}_{-0.010}$ | $13^{+12}_{-6}$ | $1.15^{+2.67}_{-0.80}$ | $5.45^{+0.31}_{-0.29}$ | $0.102^{+0.019}_{-0.011}$ |
| $6.036^{+0.015}_{-0.013}$ | $633^{+246}_{-256}$ | $0.05^{+0.01}_{-0.01}$ | | |

$N_T$ (Total): $2.02^{+2.68}_{-0.82} \times 10^{11}$ cm$^{-2}$

## 6. DISCUSSION

### 6.1. Chemical Modeling Predictions

Building on the modeling efforts of carbon-chain chemistry in Siebert et al. (2022), we utilized the nautilus v1.1 code (Ruaud et al. 2016) which has been used previously to successfully study the formation of carbon-chain molecules detected with GOTHAM data (Xue et al. 2020; McGuire et al. 2021; Shingledecker et al. 2021). The model's physical conditions are identical to those studies ($T_{gas} = T_{grain} = 10$ K, $n_{H_2} = 2 \times 10^4$ cm$^{-3}$, $A_V = 10$, and $\zeta_{CR} = 1.3 \times 10^{-17}$ s$^{-1}$; Hincelin et al. (2011)) as are the elemental abundances (Loomis et al. 2021). Based originally off of the KIDA network, our network already contained some formation routes to the $C_nH$ family from $n = 2$ to $n = 10$ and the $C_nH^-$ family from $n = 4$ to $n = 10$.

The simulated abundances are compared with those observed in TMC-1 and the machine learning predictions discussed in Section 6.2 assuming a TMC-1 hydrogen column density of $N_H = 10^{22}$ cm$^{-2}$. For utility of comparison, we adopt the same source age as discussed in Siebert et al. (2022). However, it should be noted that the simulated time of peak abundance can vary between species, with heavier species and longer carbon chains typically requiring a longer time to form. This time dependence is shown in greater detail in Appendix I.

As Figure 4 shows for the $C_nH$ family, the chemical model agrees within within a factor of 5 and the log-linear trend is generally reproduced. One of the primary production pathways of this family comes from atomic

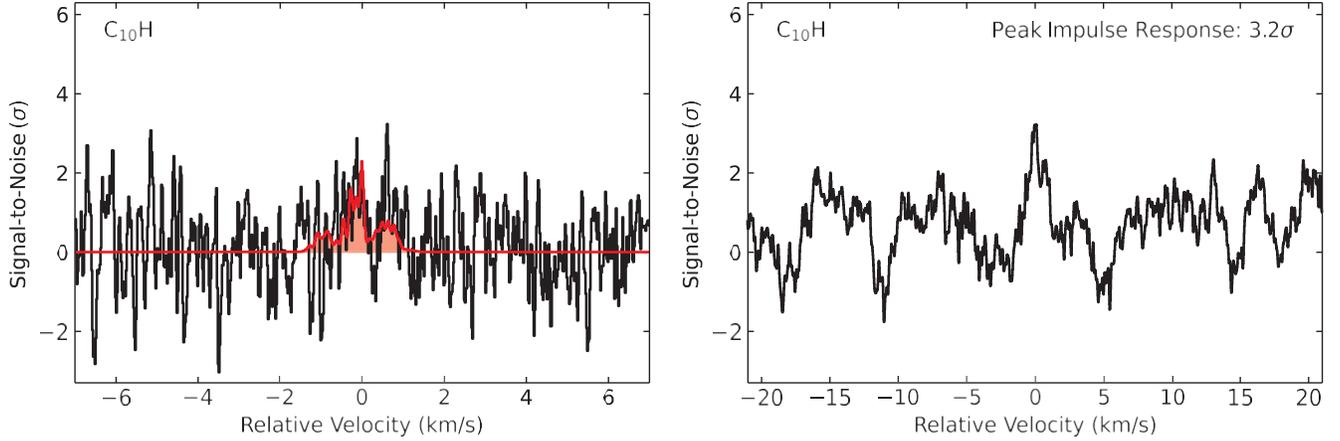

**Figure 3.** Similar to Figure 2. Velocity stacked and matched filter spectra of $C_{10}H$.

reactions of the form

$$C + C_{n-1}H_2 \rightarrow C_nH + H \quad (1)$$

The rate coefficients of these reactions were estimated by Loison et al. (2014) through extrapolation of the calculations by Chastaing, D. et al. (2001) and Chastaing et al. (2000) on C + alkenes, alkynes, dienes and diynes. The other major formation routes involve atomic reactions with related anions:

$$C + C_{n-1}H^- \rightarrow C_nH + e^- \quad (2)$$

$$H + C_n^- \rightarrow C_nH + e^- \quad (3)$$

For $C_{10}H$, these reactions were studied theoretically in Harada & Herbst (2008) and experimentally in Eichelberger et al. (2007) and Barckholtz et al. (2001). The majority of the family is primarily destroyed through electron attachment, producing their carbon-chain length analogs among the $C_nH^-$ family. These rates were also estimated by Harada & Herbst (2008).

For the carbon-chain anions, $C_nH^-$, the abundances agree within an order of magnitude for $C_6H$ and longer. While the model predictions show the monotonically decreasing abundances found for the radical species, akin to the log-linear trend seen for the majority of long carbon chains, it is not able to reproduce the enhanced abundances found in $C_6H^-$ and $C_{10}H^-$ nor the relative positive correlation between $C_nH^-$ abundance and carbon-chain length. Given the measured values, the $C_{10}H^-/C_{10}H$ column density ratio is $\sim 2.0^{+5.9}_{-1.6}$ - the highest value measured between an anion and neutral species to date. Such a high ratio is impossible for our model to reproduce, even if every electron collision with $C_{10}H$ results in $C_{10}H^-$ formation. As such, the predictions from the gas/grain chemical model illustrate that there is much still uncertain about the chemistry needed to form molecular anions in astronomical environments. The chemical network for these species originates from estimations done by Harada & Herbst (2008). The primary production route of the anions (80-90%) is through radiative electron attachment

$$C_nH + e^- \rightarrow C_nH^- + \gamma \quad (4)$$

Another minor production route is also present involving reactions between longer carbon-chain anions and atomic oxygen,

$$C_{n+1}H^- + O \rightarrow C_nH^- + CO. \quad (5)$$

There are two product channels suggested for the destruction reactions between carbon-chain anions and atomic hydrogen (Harada & Herbst 2008). We considered the pathway involving associative electron detachment,

$$C_nH^- + H \rightarrow C_nH_2 + e^-, \quad (6)$$

for $C_nH^-$ (n=4, 6, 8, and 10). In contrast, the other pathway involves fragmentation and has only been considered for $C_{10}H^-$,

$$C_{10}H^- + H \rightarrow C_8H^- + C_2H. \quad (7)$$

Harada & Herbst (2008) originally estimated all rates for route 6 to be $1.0 \times 10^{-9}$ cm$^3$ s$^{-1}$. This was also the rate estimated for route 7. In order to keep the total $C_{10}H^-$ + H rate consistent with the other analogous anion reactions in route 6, we modified the rates of both $C_{10}H^-$ + H reactions from routes 6 and 7 from $1.0 \times 10^{-9}$ cm$^3$ s$^{-1}$ to $5.0 \times 10^{-10}$ cm$^3$ s$^{-1}$. This resulted in a factor of $\sim 2$ increase in the abundance of $C_{10}H$ relative to simulations performed with the original rates.

There are limited experimental constraints on these pathways and the corresponding rate coefficients. In particular, the major route 4 is very challenging to measure experimentally because of twofold reasons. Firstly, anions are difficult to produce in a stable quantity. Secondly, it is difficult to observe the photoemission process due to competing collisional stabilization. In addition, anion-neutral reactions, for example, might also contribute to the $C_nH^-$ 's formation, but these need to be investigated further. Furthermore, while the current model mainly focuses on the gas chemistry of the $C_nH$ species, electron attachment processes might occur on grains, allowing the super excited anion intermediate to be stabilized efficiently rather than being dependent on radiative processes in the gas phase. Our understanding of molecular anions and carbon chains can be improved by constraining their grain chemistry.

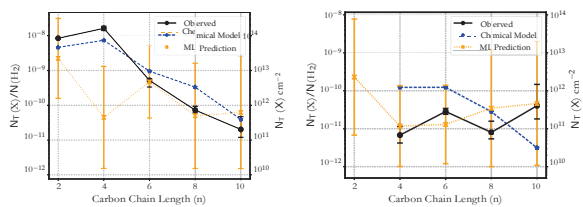

**Figure 4.** The abundance and column density of the $C_nH$ (left) and $C_nH^-$ (right) families as observed by the GBT (black, solid, circles), simulated by nautilus (blue, dashed, stars) at a model time of $t = 2.5 \times 10^5$ years, and predicted by machine learning (orange, dotted, $\times$s). The observed column density of $C_2H$ is taken from Pratap et al. (1997) with the Five College Radio Astronomical Observatory (FCRAO) 14m antenna .

### 6.2. Machine Learning Predictions

The column densities of both the observed and unobserved parent and anion pairs were predicted using the trained supervised learning regressors presented in Lee et al. (2021b). For brevity, we only explain the relevant methodological details here.

Starting from SMILES (Simplified Molecular-Input Line-Entry System) (Weininger 1988; O'Boyle 2012) representations of molecules, we rely on a pretrained MOL2VEC embedding model (Jaeger et al. 2018) to transform each species into corresponding molecular vectors. These vectors form a high dimensional representation that captures chemical (e.g. charge state, bonding patterns) and structural (e.g. aromaticity) properties into a compact form usable by even simple regressors for property prediction; in this case, and that of Lee et al. (2021b), to predict column densities of unseen molecules. In contrast to conventional chemical modeling, our machine learning approach only requires molecular structure as input, and prior knowledge of the chemical and physical properties of the source are captured implicitly based on what molecules have been observed. Thus, when conditioned on an astrophysical environment, the trained model predictions can be interpreted as a baseline for column densities based *solely* on manifold distances from new molecules to those already observed.

The column density predictions from the machine learning approach are given in Table 4, combining two types of models as described in Lee et al. (2021b) using regression modules from scikit-learn (Pedregosa et al. 2012). The magnitude of the column densities were predicted using the gradient boosting regressor (GBR), given its high accuracy in many regression tasks. We also use a Gaussian process regressor (GPR) to estimate the uncertainty in the column density predictions. Aside from $C_4H$, our machine learning predictions match the mean of the observed values well within an order of magnitude, and are captured within our observational and model uncertainties.

In case of $C_4H$, the nearly degenerate low-lying electronic states (Senent & Hochlaf 2010) cause a discrepancy in the dipole moment trend that is observed across the other carbon chains investigated in this work. The average dipole resulting from a mixed state is not well understood, as discussed in Gratier et al. (2016). We attribute the low column density prediction for $C_4H$ to limitations in the ML predictions, as they are based purely on chemical similarity without dynamical/electronic effects being considered, and can be considered as a baseline; their main strength is generalizing across families of molecules to unseen species in a fast and data-driven manner.

The inherent value of the machine learning predictions is two-fold: the capability to provide baseline expectation for the $C_{10}H/C_{10}H^-$ ratio and a straightforward means to predict the abundance for the next in series carbon chains, $C_{12}H/C_{12}H^-$ . The predictions from the machine learning approach are able to better match both the radical and anion species detected towards TMC-1 compared to the chemical formation model, except for the predictions of $C_4H$ due to electronic effects (as discussed above) and expected model uncertainty for species where representative inventory and knowledge is sparse or lacking. The machine learning approach also comes closer to predicting the observed $C_{10}H^-$ /$C_{10}H$ column density ratio compared to the chemical formation model. As such, it is now possible to make a prediction for the other non-detected carbon chains species, namely $C_{12}H$ and $C_{12}H^-$ and the smaller molecular anion $CCH^-$. For comparison, the machine learning model prediction for CCH of $2.2 \times 10^{13}$ cm$^{-2}$ is within a factor of 3 of the previously measured value of $7.2 \times 10^{13}$ cm$^{-2}$ found by Pratap et al. (1997). For $CCH^-$ , chem-

**Table 4.** Column densities used to train the machine learning algorithm and the predicted column density outputs

| Molecule | $N_T$ | $N_T$ + Std dev. | $N_T$ - Std dev. |
|----------|-------|------------------|------------------|
| | $(10^{11} \mathrm{cm}^{-2})$ | $(10^{11} \mathrm{cm}^{-2})$ | $(10^{11} \mathrm{cm}^{-2})$ |
| $C_2H$ | 223.0620 | 3138.7216 | 15.8525 |
| $C_4H$ | 4.4814 | 131.2213 | 0.1530 |
| $C_6H$ | 46.8980 | 512.2110 | 4.2940 |
| $C_8H$ | 4.9852 | 161.8065 | 0.1536 |
| $C_{10}H$ | 6.2294 | 255.1943 | 0.1521 |
| $C_{12}H$ | 7.1877 | 364.4581 | 0.1417 |
| $C_2H^-$ | 22.9619 | 777.7198 | 0.6779 |
| $C_4H^-$ | 1.1724 | 13.5263 | 0.1016 |
| $C_6H^-$ | 1.2983 | 14.0185 | 0.1203 |
| $C_8H^-$ | 3.4520 | 118.5992 | 0.1005 |
| $C_{10}H^-$ | 4.6145 | 196.5918 | 0.1083 |
| $C_{12}H^-$ | 7.0534 | 395.2162 | 0.1259 |

The column densities predictions using gradient boosting regression. The standard deviations for the predictions were calculated using Gaussian Process Regression.

ical model predictions are not as clear: earlier chemical model predictions suggests CCH$^-$ should have a lower abundance than the larger polyyne anions due to its greatly reduced radiative electron attachment rate (Herbst & Osamura 2008; Cordiner et al. 2008). Yet, the machine learning model prediction of $2.3 \times 10^{12}$ cm$^{-2}$ suggests that CCH$^-$ may be abundant enough to be detected in sources such as TMC-1. However, the results from Agúndez et al. (2008) reported an upper limit for CCH$^-$ of $< 2.2 \times 10^{11}$ cm$^{-2}$ towards TMC-1. As such, these disparate set of predictions compared to the reported upper limit again shows our limited understanding of the formation of the smaller CCH$^-$ anion. For the larger species $C_{12}H$ and $C_{12}H^-$, the predictions of $7.1 \times 10^{11}$ cm$^{-2}$ and $7.0 \times 10^{11}$ cm$^{-2}$, respectively, also suggest they may be detected in TMC-1 once the spectroscopy of these species (including calculated or measured dipole moments and partition functions), specifically $C_{12}H^-$ are fully characterized. The spectroscopy of $C_{12}H$ is reported in Gottlieb et al. (1998a) and can be used to guide an astronomical search.

## 7. CONCLUSIONS

Using the GOTHAM data, we report the first astronomical detection of the $C_{10}H^-$ anion toward the dark cloud TMC-1 with the GBT. In fact, the astronomical observations provided the necessary data to the refinement of the spectroscopic parameters of $C_{10}H^-$ to enable the first astronomical detection. These new parameters will be essential for future studies of $C_{10}H^-$ in the laboratory. From the velocity stacked data and the matched filter response, $C_{10}H^-$ is detected at $>9\sigma$ confidence interval. In addition, there is evidence for several individual lines of $C_{10}H^-$ in the GOTHAM data though none above the current noise level of the survey beyond the $>3\sigma$ limit. In this case, 3 independent velocity components were detected and a total $C_{10}H^-$ column density of $4.04^{+0.67}_{-0.23} \times 10^{11}$ cm$^{-2}$ was determined.

A dedicated search for the $C_{10}H$ radical was also conducted towards TMC-1. The stacked molecular emission of $C_{10}H$ was detected at a $\sim3.2\sigma$ confidence interval. As such, the presence of this molecule is currently considered tentative. In addition, no individual spectral lines from this species were detected. The measured physical values determined from the $C_{10}H$ fits include 4 independent velocity components and a total $C_{10}H$ column density of $2.02^{+2.68}_{-0.82} \times 10^{11}$ cm$^{-2}$ was reported. Given the measured values, the $C_{10}H^-$ /$C_{10}H$ column density ratio is $\sim2.0^{+5.9}_{-1.6}$, the highest value measured between an anion and neutral species to date. Such a high ratio is at odds with current theories for interstellar anion chemistry.

The full GOTHAM dataset was also used to better characterize the physical parameters including column density [$N_T$], excitation temperature [$T_{ex}$], linewidth [$\Delta V$], and source size [$''$] for the more abundant $C_4H$, $C_4H^-$, $C_6H$, $C_6H^-$, $C_8H$, and $C_8H^-$ molecules. These data were compared to predicted abundances from both a gas/grain chemical formation model and from a machine learning analysis. For the radical species, both models reproduce the measured abundances found from the survey better than an order of magnitude (except for the $C_4H$ predicted abundance from the machine learning analysis). However, the machine learning analysis matches the detected anion abundances much better than the gas/grain chemical model suggesting that the understanding of the formation chemistry of molecular anions is still highly questionable. Finally, using the machine learning analysis, it is possible to make a prediction for the larger species, namely $C_{12}H$ and $C_{12}H^-$ and the smaller molecular anion CCH$^-$. For CCH$^-$, a model prediction of $2.3 \times 10^{12}$ cm$^{-2}$ shows that predicted column density of CCH$^-$ is in stark contrast to the reported upper limit of $< 2.3 \times 10^{12}$ cm$^{-2}$. However, for the larger species $C_{12}H$ and $C_{12}H^-$, the predictions of $7.1 \times 10^{11}$ cm$^{-2}$ and $7.0 \times 10^{11}$ cm$^{-2}$, respectively, suggest they may be detected in TMC-1 once the spectroscopy of these species are fully characterized for an astronomical search.

The authors would like to thank the anonymous referees who provided valuable insight and suggestions that greatly improved the manuscript. We gratefully acknowledge support from NSF grants AST-1908576 and AST-2205126. I.R.C. acknowledges support from the University of British Columbia and the Natural Sciences and Engineering Research Council of Canada (NSERC). E. H. thanks the National Science Foundation (US) for support of his research program in astrochemistry through grant AST 19-06489. M.A.C. and S.B.C. were supported by the Goddard Center for Astrobiology. H.G. acknowledges support from the National Science Foundation for participation in this work as part of his independent research and development plan. Any opinions, findings, and conclusions expressed in this material are those of the authors and do not necessarily reflect the views of the National Science Foundation. The National Radio Astronomy Observatory is a facility of the National Science Foundation operated under cooperative agreement by Associated Universities, Inc. The Green Bank Observatory is a facility of the National Science Foundation operated under cooperative agreement by Associated Universities, Inc.

# APPENDIX

## A. C$_{10}$H/C$_{10}$H$^-$ ANALYSIS

Fig. A1 shows individual lines of C$_{10}$H$^-$ covered in the GOTHAM data through 25 GHz. The data and simulations are displayed at a smoothed resolution of 5.6 kHz (versus the 1.4 kHz native resolution of the GOTHAM observations) to show the (very weak) potential signal from these individual lines particularly between 13.1–19.2 GHz. Lines up to 30 GHz were included in the analysis (and are included in the catalog provided in the Supplementary Information), but are not shown here as the noise level of that data is much higher than the predicted line intensities.

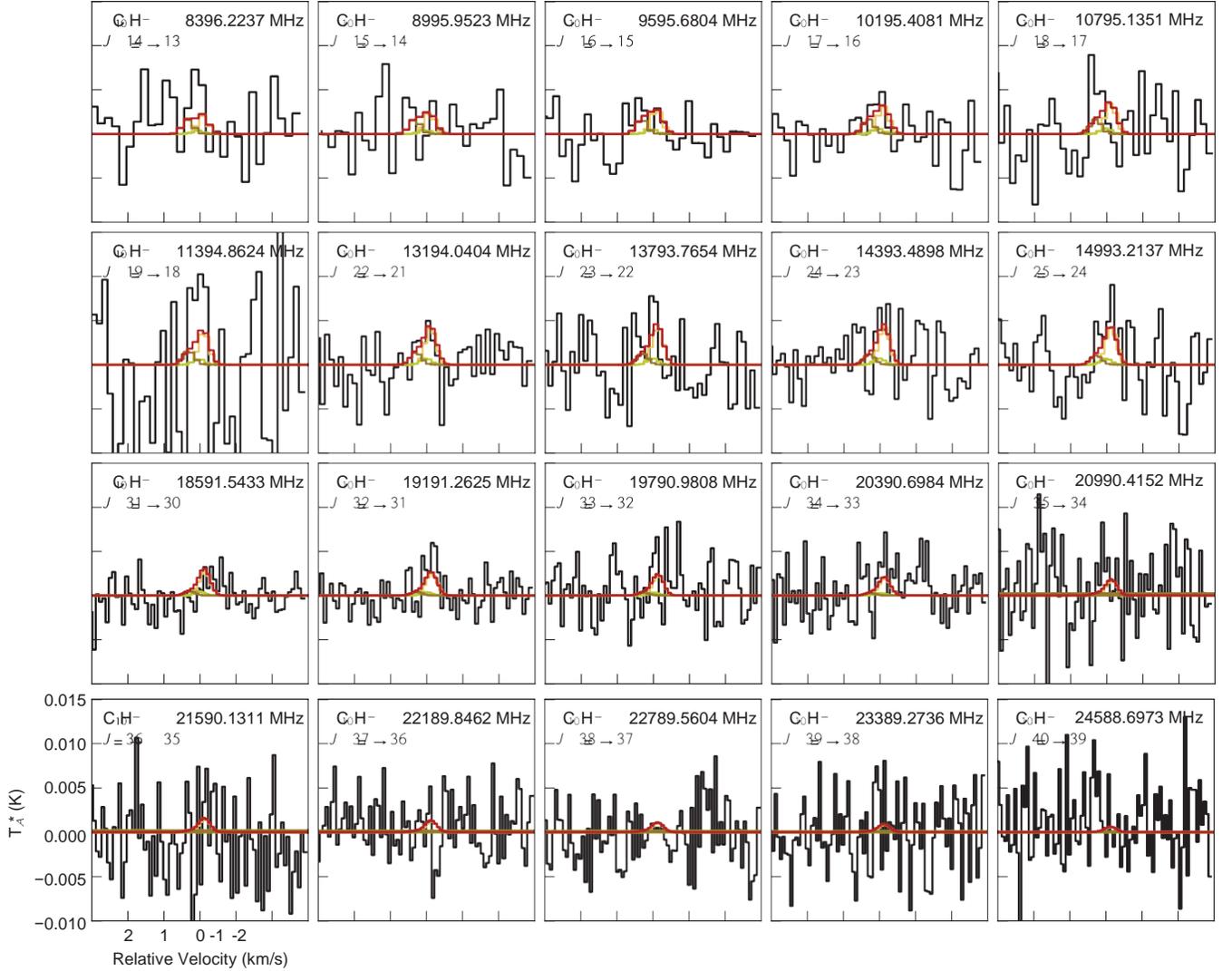

**Figure A1.** Individual lines of C$_{10}$H$^-$ covered in the GOTHAM observations through 25 GHz (black). Both the spectra and the simulations have been smoothed to a resolution of 5.6 kHz (from the native 1.4 kHz full resolution) to better show potential features. The spectra for lines falling at higher frequencies are substantially noisier and have been omitted from the plot to reduce the number of panels. Simulations of C$_{10}$H$^-$ emission using the parameters given in Table 2 are shown in colors, with the total simulation in red. The quantum numbers for each transition are given in the upper left of each panel, and the central frequency of the window (in the sky frame) is given in the top right. Each window is 6.0 km s$^{-1}$ in total width.

Figs. A2 & A3 show the corner plots from the MCMC analysis of C$_{10}$H$^-$ and C$_{10}$H, respectively.

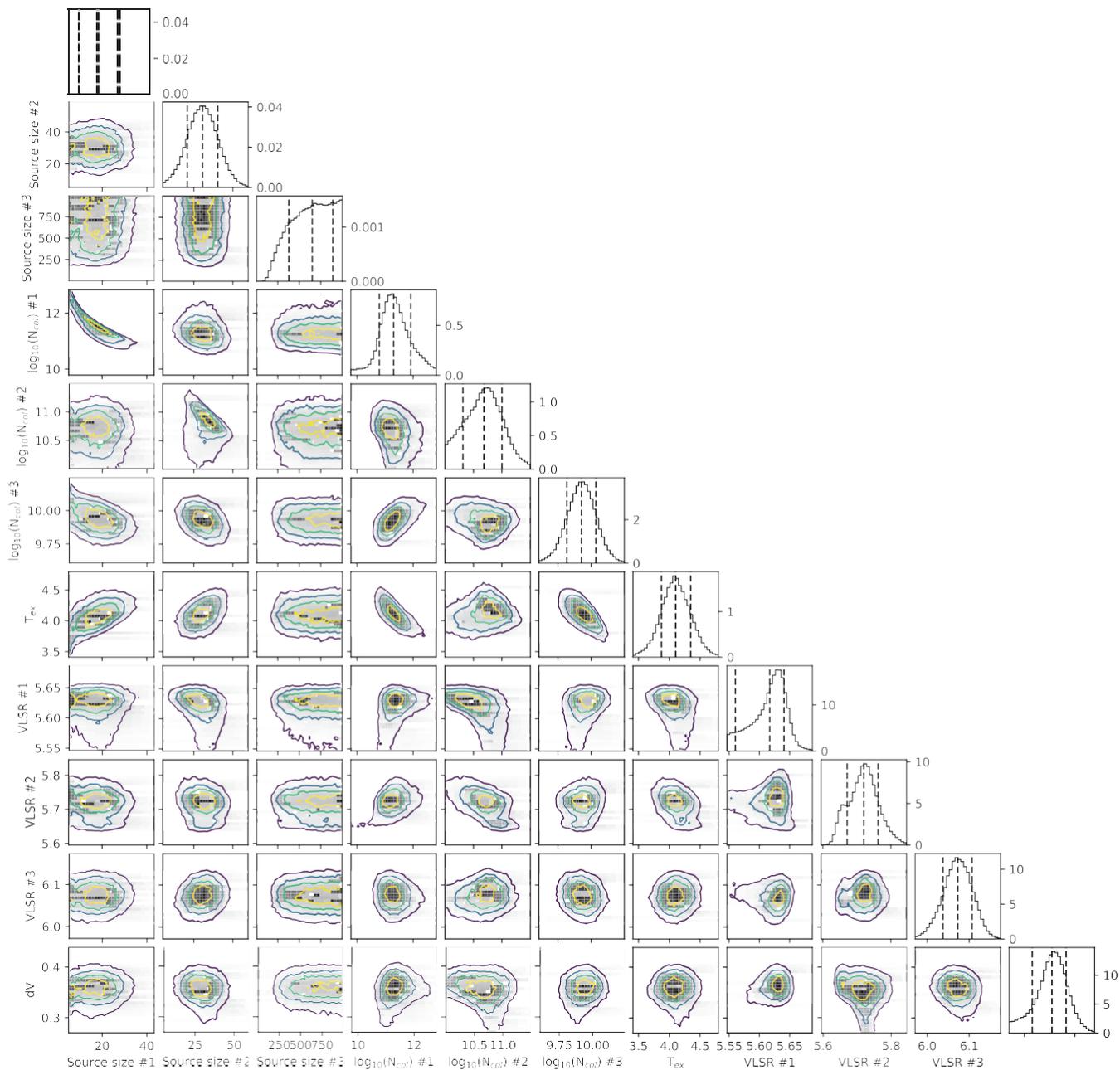

**Figure A2.** Corner plot for $C_{10}H^-$ showing parameter covariances and marginalized posterior distributions for the $C_{10}H^-$ MCMC fit. $16^{th}$, $50^{th}$, and $84^{th}$ confidence intervals (corresponding to $\pm 1$ sigma for a Gaussian posterior distribution) are shown as vertical lines.

### A.1. *Jacknife Analysis*

To further ensure that the signal attributed to $C_{10}H^-$ is molecular in origin, we performed a jacknife analysis similar to that described in detail in McGuire et al. (2021). The catalog for $C_{10}H^-$ was divided in half, with every other line assigned to one of two different catalogs. A spectral stack and matched filter was then performed on each catalog separately. Assuming the signal is indeed molecular and coming from $C_{10}H^-$, the resulting impulse response functions should, when added in quadrature, should closely reproduce the signal from the full catalog. The results of the analysis are shown in Fig. A4, and we find impulse response functions of $5.3\sigma$ and $7.8\sigma$. When added in quadrature, this results in a total of $9.2\sigma$, in very good agreement with the value determined from the entire catalog and shown in Fig. 2 of

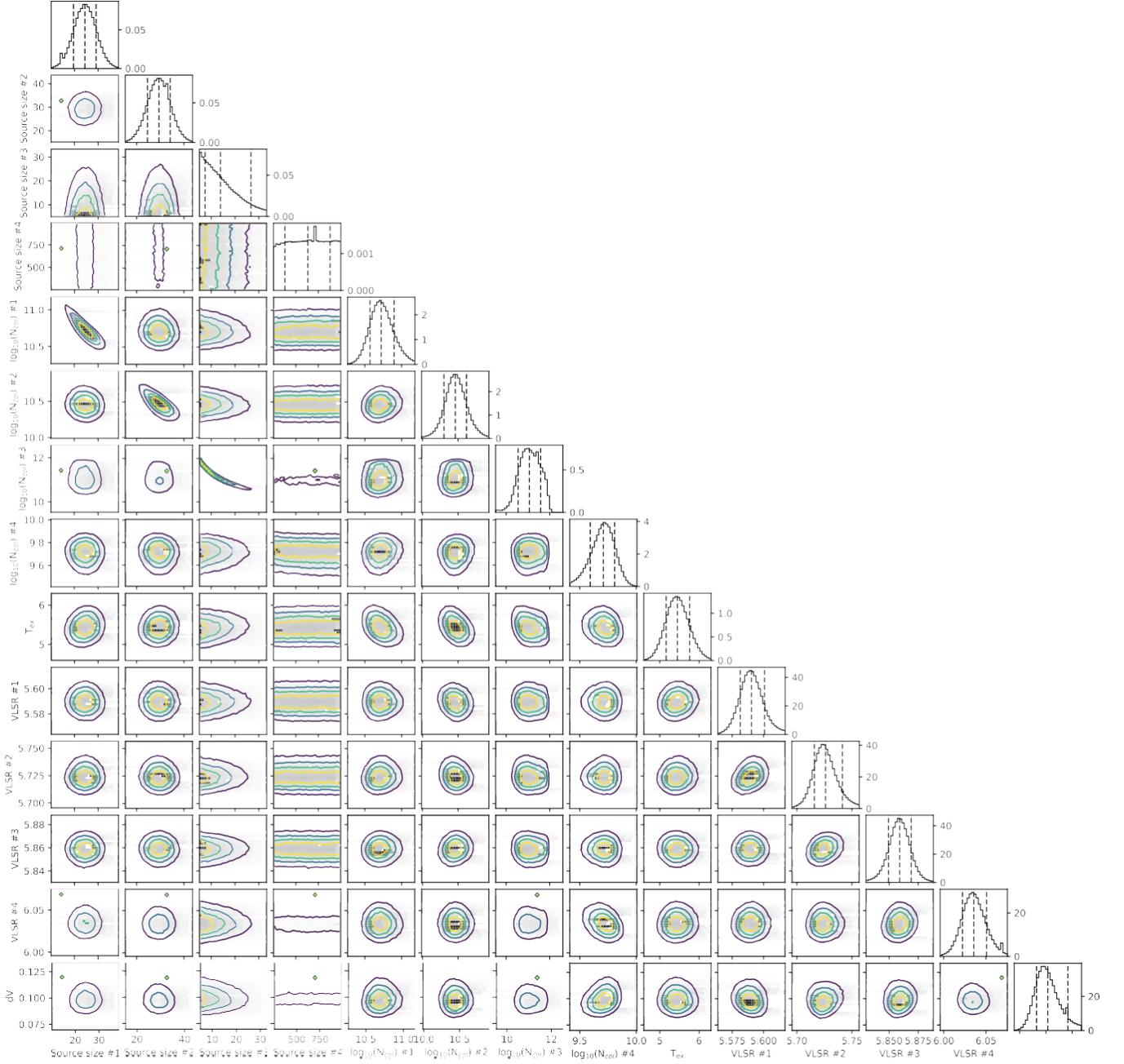

**Figure A3.** Corner plot for C$_{10}$H showing parameter covariances and marginalized posterior distributions for the C$_{10}$H MCMC fit. 16$^{th}$, 50$^{th}$, and 84$^{th}$ confidence intervals (corresponding to $\pm 1$ sigma for a Gaussian posterior distribution) are shown as vertical lines.

9.3$\sigma$. As discussed in McGuire et al. (2021), the small mismatch is almost certainly due to minor contributions of red noise in the data at the level of, in this case, 0.1$\sigma$.

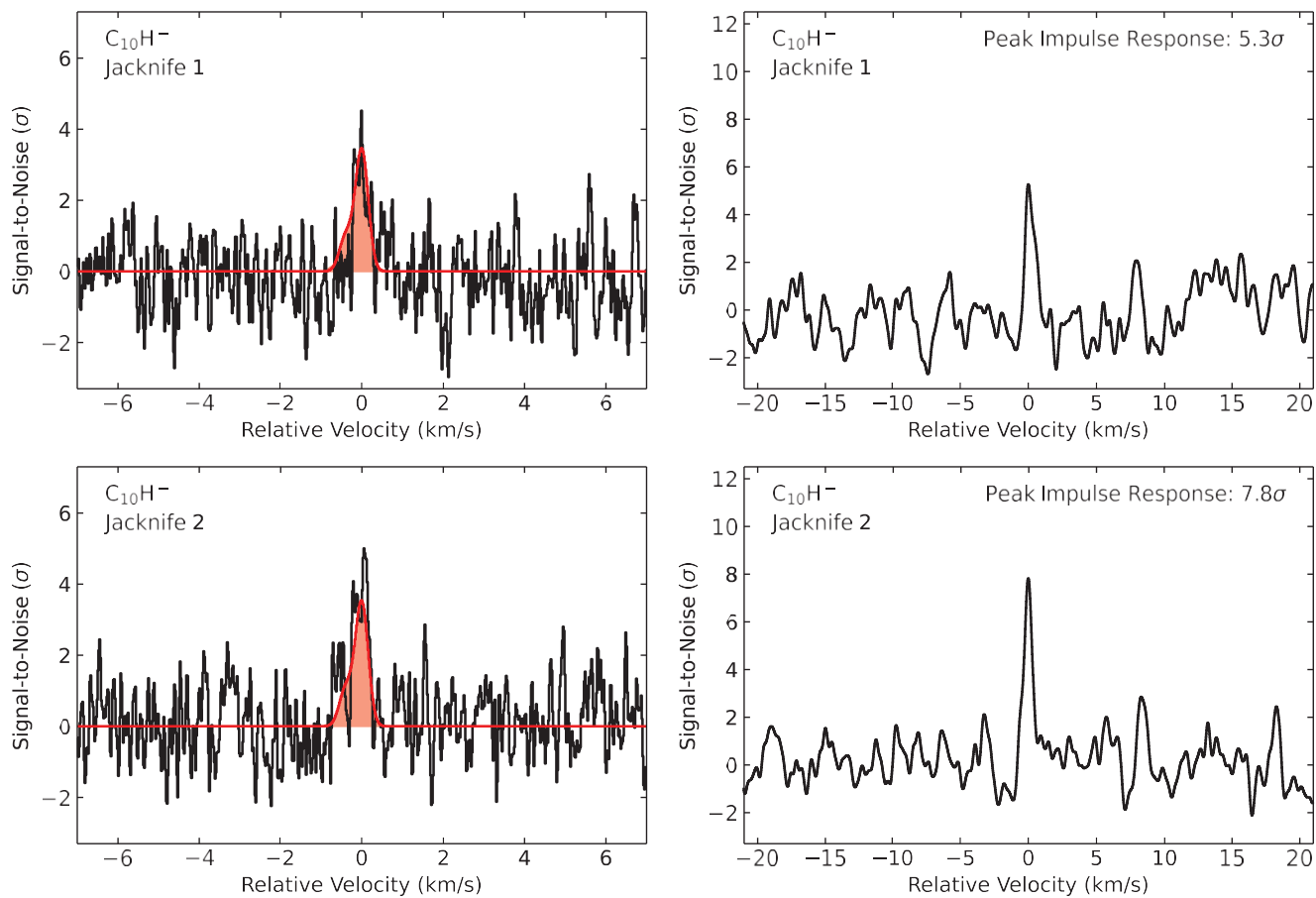

**Figure A4.** Velocity stacked and matched filter spectra of $C_{10}H^-$ from a jacknife analysis where the full catalog was split into two. The intensity scales are the signal-to-noise ratios (SNR) of the response functions when centered at a given velocity. The "zero" velocity corresponds to the channel with the highest intensity to account for blended spectroscopic transitions and variations in velocity component source sizes. (*Left*) The stacked spectra from the GOTHAM DR4 data are displayed in black, overlaid with the expected line profile in red from our MCMC fit to the data. The signal-to-noise ratio is on a per-channel basis. (*Right*) Matched filter response obtained from cross-correlating the simulated and observed velocity stacks in the left panel; value annotated corresponds to the peak impulse response of the matched filter.

## B.  $C_4H$ ANALYSIS

The best-fit parameters from the MCMC fit for $C_4H$ are shown in Table B1. The corner plot from the analysis is shown in Fig. B1. Figs. B2−B4 show the individual lines of $C_4H$ detected in the GOTHAM observations.

**Table B1.** $C_4H$ Values

| $v_{lsr}$ (km s$^{-1}$) | Size ($''$) | $N_T^\dagger$ ($10^{13}$cm$^{-2}$) | $T_{ex}$ (K) | $\Delta V$ (km s$^{-1}$) |
|---|---|---|---|---|
| $5.657^{+0.008}_{0.009}$ | $570^{+289}_{-300}$ | $6.95^{+0.85}_{-0.94}$ | | |
| $5.754^{+0.029}_{0.030}$ | $29^{+5}_{-5}$ | $6.52^{+2.24}$ | $5.10^{+0.46}_{-0.42}$ | $0.178^{+0.013}_{-0.010}$ |
| $5.864^{+0.045}_{0.032}$ | $543^{+311}_{-332}$ | $1.42^{+0.75}_{-0.68}$ | | |
| $6.047^{+0.019}_{0.025}$ | $453^{+369}_{-292}$ | $1.33^{+0.30}_{-0.26}$ | | |

$N_T$(Total)$^{\dagger\dagger}$: $1.62^{+0.25}_{-0.22} \times 10^{14}$ cm$^{-2}$

Note – The quoted uncertainties represent the $16^{th}$ and $84^{th}$ percentile ($1\sigma$ for a Gaussian distribution) uncertainties.
$^\dagger$Column density values are highly covariant with the derived source sizes.
$^{\dagger\dagger}$Uncertainties derived by adding the uncertainties of the individual components in quadrature.

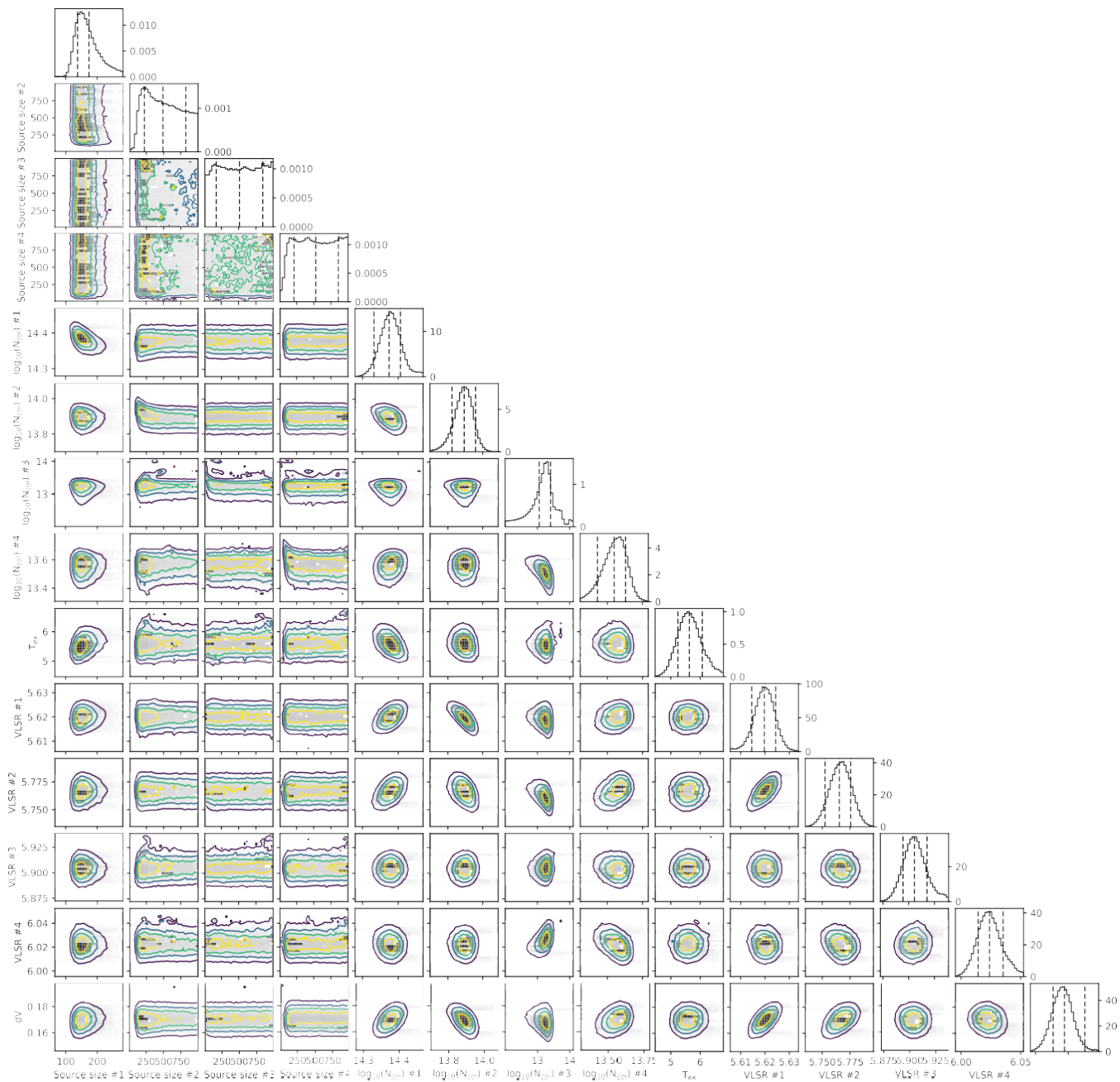

**Figure B1.** Corner plot for C$_4$H showing parameter covariances and marginalized posterior distributions for the C$_4$H MCMC fit. 16$^{th}$, 50$^{th}$, and 84$^{th}$ confidence intervals (corresponding to ±1 sigma for a Gaussian posterior distribution) are shown as vertical lines.

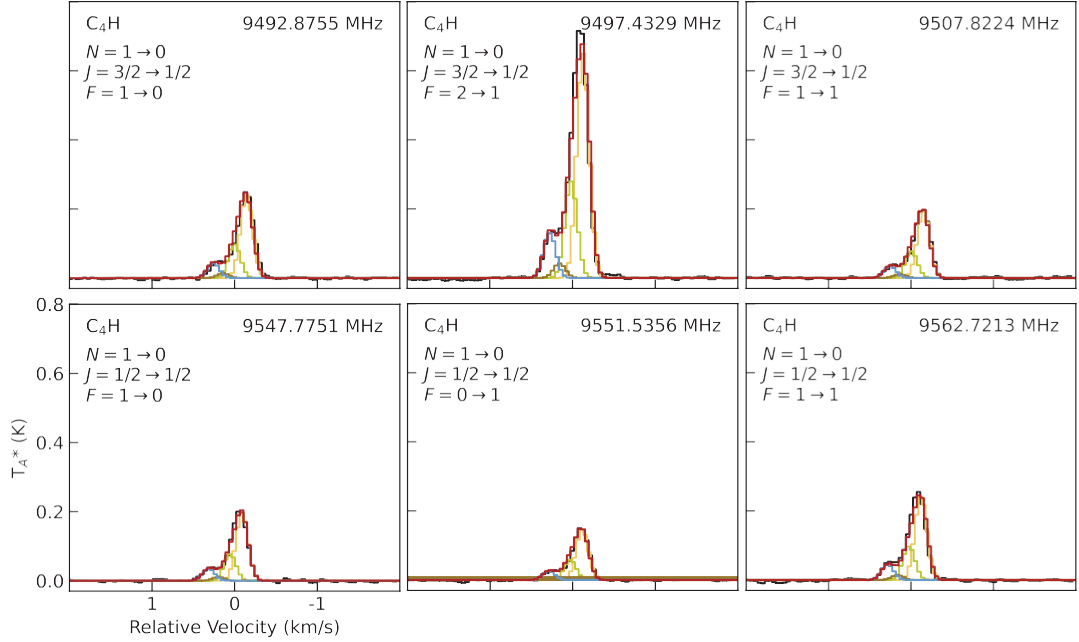

**Figure B2.** Individual lines of $C_4H$ detected in the GOTHAM observations (black). Simulations of $C_4H$ emission using the parameters given in Table B1 are shown in colors, with the total simulation in red. The quantum numbers for each transition are given in the upper left of each panel, and the central frequency of the window (in the sky frame) is given in the top right. Each window is 4.0 km s$^{-1}$ in total width.

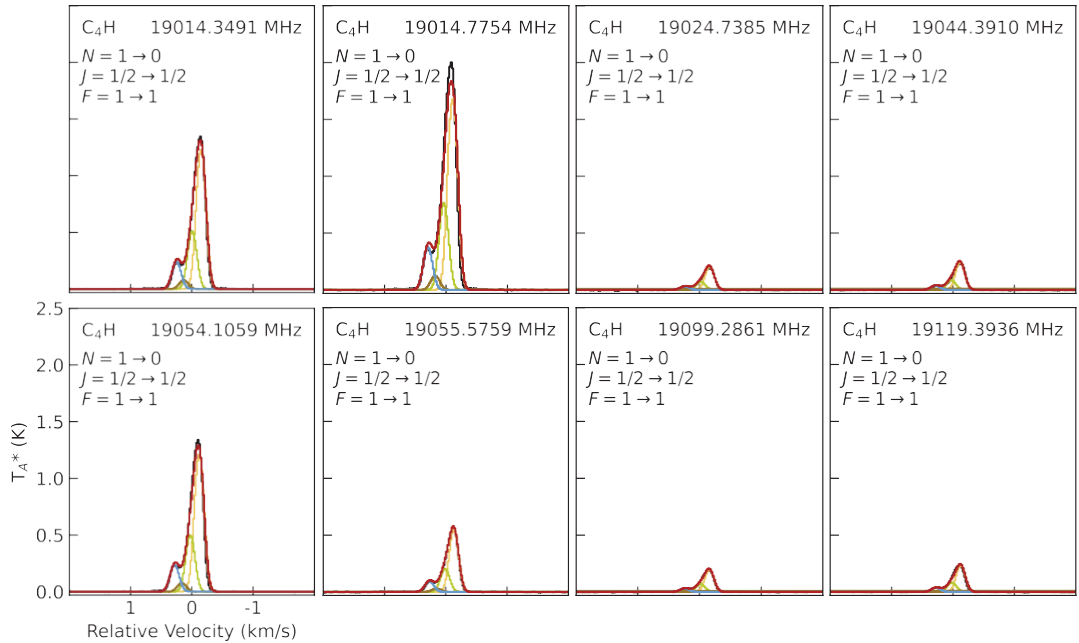

**Figure B3.** Individual lines of $C_4H$ detected in the GOTHAM observations (black). Simulations of $C_4H$ emission using the parameters given in Table B1 are shown in colors, with the total simulation in red. The quantum numbers for each transition are given in the upper left of each panel, and the central frequency of the window (in the sky frame) is given in the top right. Each window is 4.0 km s$^{-1}$ in total width.

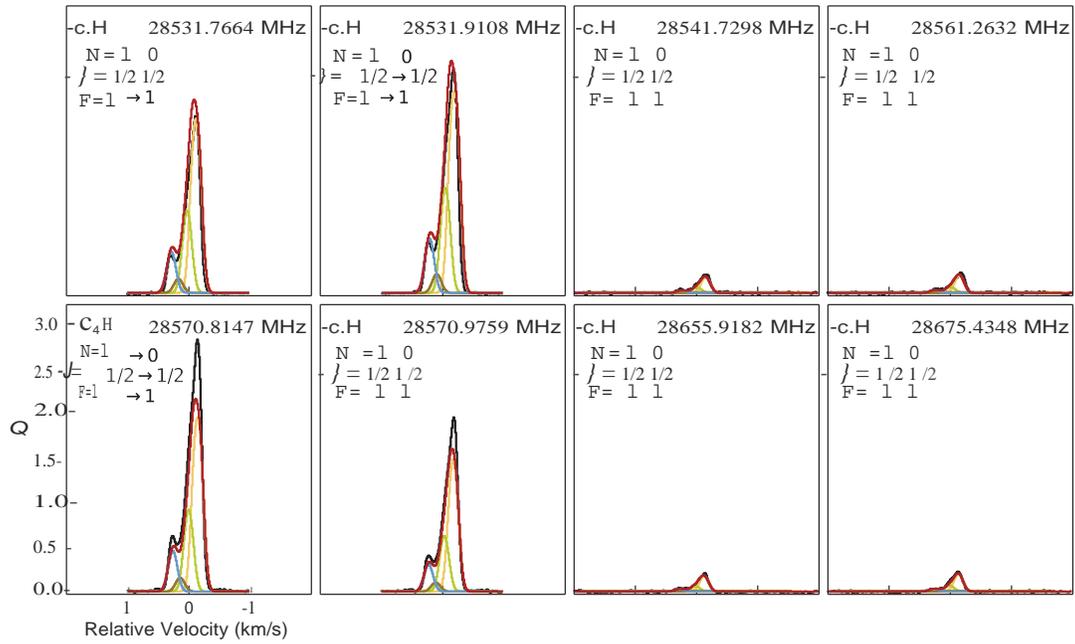

**Figure B4.** Individual lines of C$_4$H detected in the GOTHAM observations (black). Simulations of C$_4$H emission using the parameters given in Table B1 are shown in colors, with the total simulation in red. The quantum numbers for each transition are given in the upper left of each panel, and the central frequency of the window (in the sky frame) is given in the top right. Each window is 4.0 km s$^{-1}$ in total width.

## C. C$_4$H$^-$ ANALYSIS

The best-fit parameters from the MCMC fit for C$_4$H$^-$ are shown in Table C1. The corner plot from the analysis is shown in Fig. C1. Fig. C2 shows the individual lines of C$_4$H$^-$ detected in the GOTHAM observations.

**Table C1.** C$_4$H$^-$ Values

| $v_{lsr}$ (km s$^{-1}$) | Size ($''$) | $N_T$ ($10^{10}$ cm$^{-2}$) | $T_{ex}$ (K) | $\Delta V$ (km s$^{-1}$) |
|---|---|---|---|---|
| $5.652^{+0.013}_{0.032}$ | $40^{+4}_{-4}$ | $2.51^{+0.62}_{1.07}$ | | |
| $5.740^{+0.047}_{0.034}$ | $31^{+4}_{-4}$ | $0.95^{+1.32}$ | | |
| $5.916^{+0.040}_{0.047}$ | $29^{+5}_{-5}$ | $0.44^{+0.62}_{-0.39}$ | $5.13^{+0.48}_{-0.47}$ | $0.181^{+0.014}_{-0.018}$ |
| $6.018^{+0.058}_{0.035}$ | $10^{+4}_{-3}$ | $2.89^{+4.41}_{-2.22}$ | | |

$N_T$ (Total): $6.79^{+4.68}_{-2.59} \times 10^{10}$ cm$^{-2}$

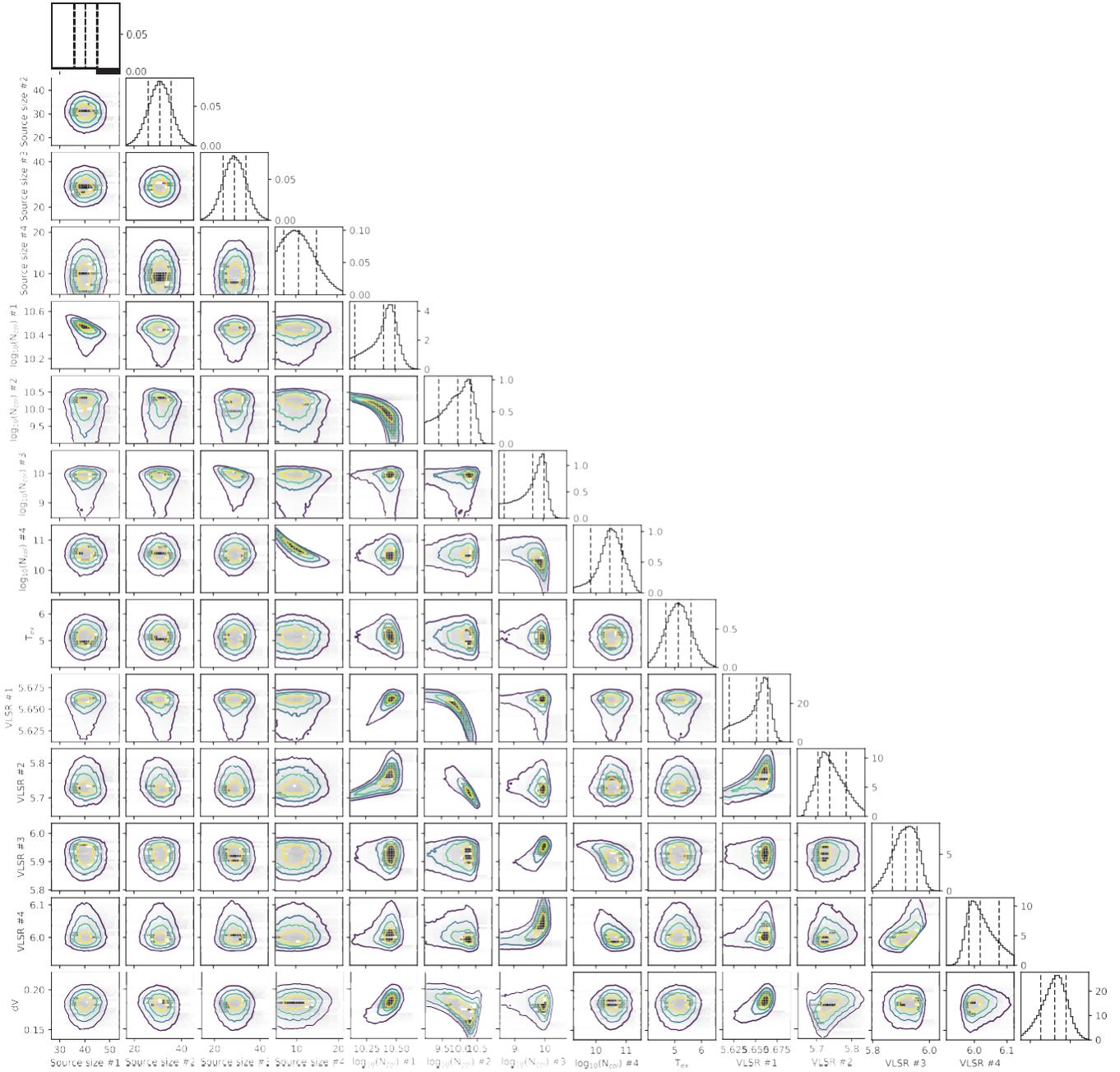

**Figure C1.** Corner plot for C$_4$H$^-$ showing parameter covariances and marginalized posterior distributions for the C$_4$H$^-$ MCMC fit. $16^{th}$, $50^{th}$, and $84^{th}$ confidence intervals (corresponding to $\pm 1$ sigma for a Gaussian posterior distribution) are shown as vertical lines.

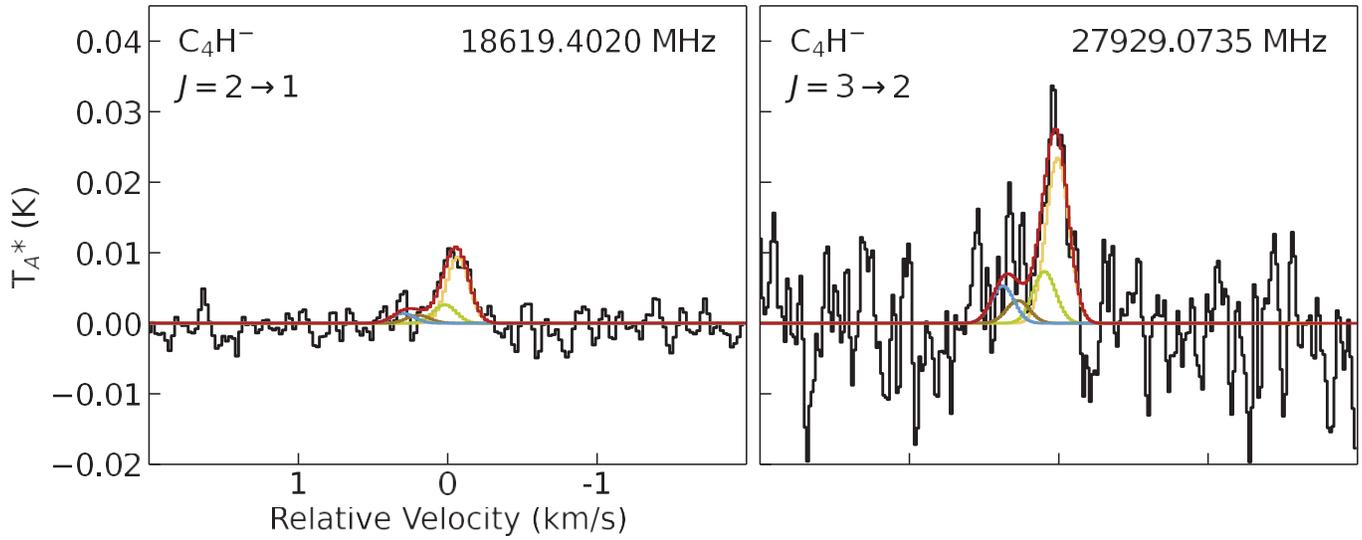

**Figure C2.** Individual lines of C$_4$H$^-$ detected in the GOTHAM observations (black). Simulations of C$_4$H$^-$ emission using the parameters given in Table C1 are shown in colors, with the total simulation in red. The quantum numbers for each transition are given in the upper left of each panel, and the central frequency of the window (in the sky frame) is given in the top right. Each window is 4.0 km s$^{-1}$ in total width.

## D. C$_6$H ANALYSIS

The best-fit parameters from the MCMC fit for C$_6$H are shown in Table D1. The corner plot from the analysis is shown in Fig. D1. Figs. D2 and D3 show the individual lines of C$_6$H detected in the GOTHAM observations.

**Table D1.** C$_6$H Values

| $v_{lsr}$ (km s$^{-1}$) | Size ($''$) | $N_T$ ($10^{12}$cm$^{-2}$) | $T_{ex}$ (K) | $\Delta V$ (km s$^{-1}$) |
|---|---|---|---|---|
| $5.600^{+0.012}_{-0.005}$ | $356^{+99}_{-110}$ | $1.15^{+0.15}_{-0.07}$ | | |
| $5.744^{+0.020}_{-0.008}$ | $35^{+55}_{-5}$ | $3.06^{+0.58}_{-1.78}$ | $5.12^{+0.13}_{-0.06}$ | $0.154^{+0.025}_{-0.008}$ |
| $5.882^{+0.021}_{-0.019}$ | $339^{+111}_{-136}$ | $0.50^{+0.08}_{-0.42}$ | | |
| $6.016^{+0.014}_{-0.023}$ | $284^{+144}_{-121}$ | $0.46^{+0.14}_{-0.07}$ | | |

$N_T$(Total): $5.17^{+0.62}_{-1.83} \times 10^{12}$ cm$^{-2}$

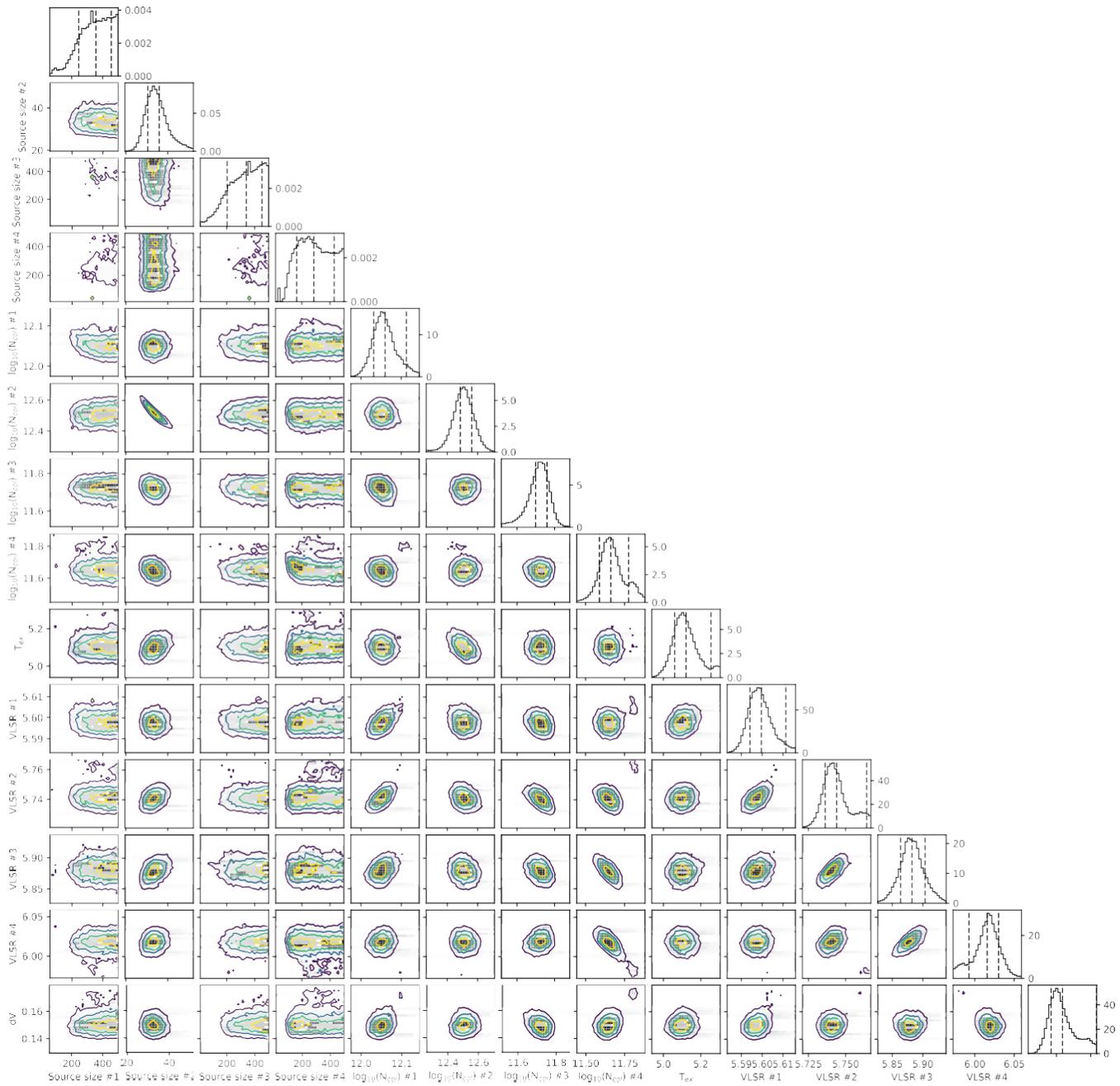

**Figure D1.** Corner plot for $C_6H$ showing parameter covariances and marginalized posterior distributions for the $C_6H$ MCMC fit. $16^{th}$, $50^{th}$, and $84^{th}$ confidence intervals (corresponding to $\pm 1$ sigma for a Gaussian posterior distribution) are shown as vertical lines.

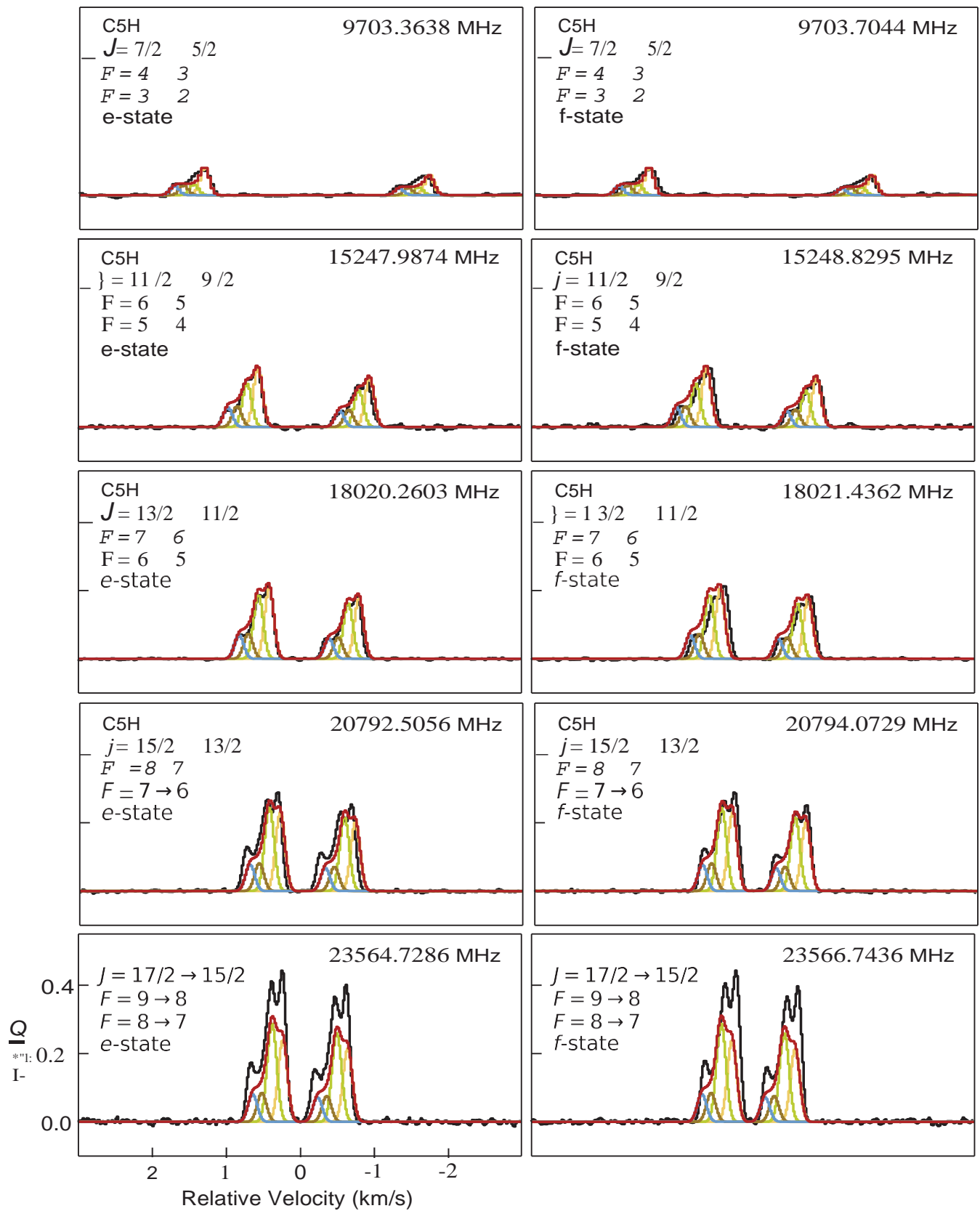

**Figure D2.** Individual lines of $C_6H$ detected in the GOTHAM observations (black). Simulations of $C_6H$ emission using the parameters given in Table D1 are shown in colors, with the total simulation in red. The quantum numbers for each transition are given in the upper left of each panel, and the central frequency of the window (in the sky frame) is given in the top right. Each window is 6.0 km s$^{-1}$ in total width.

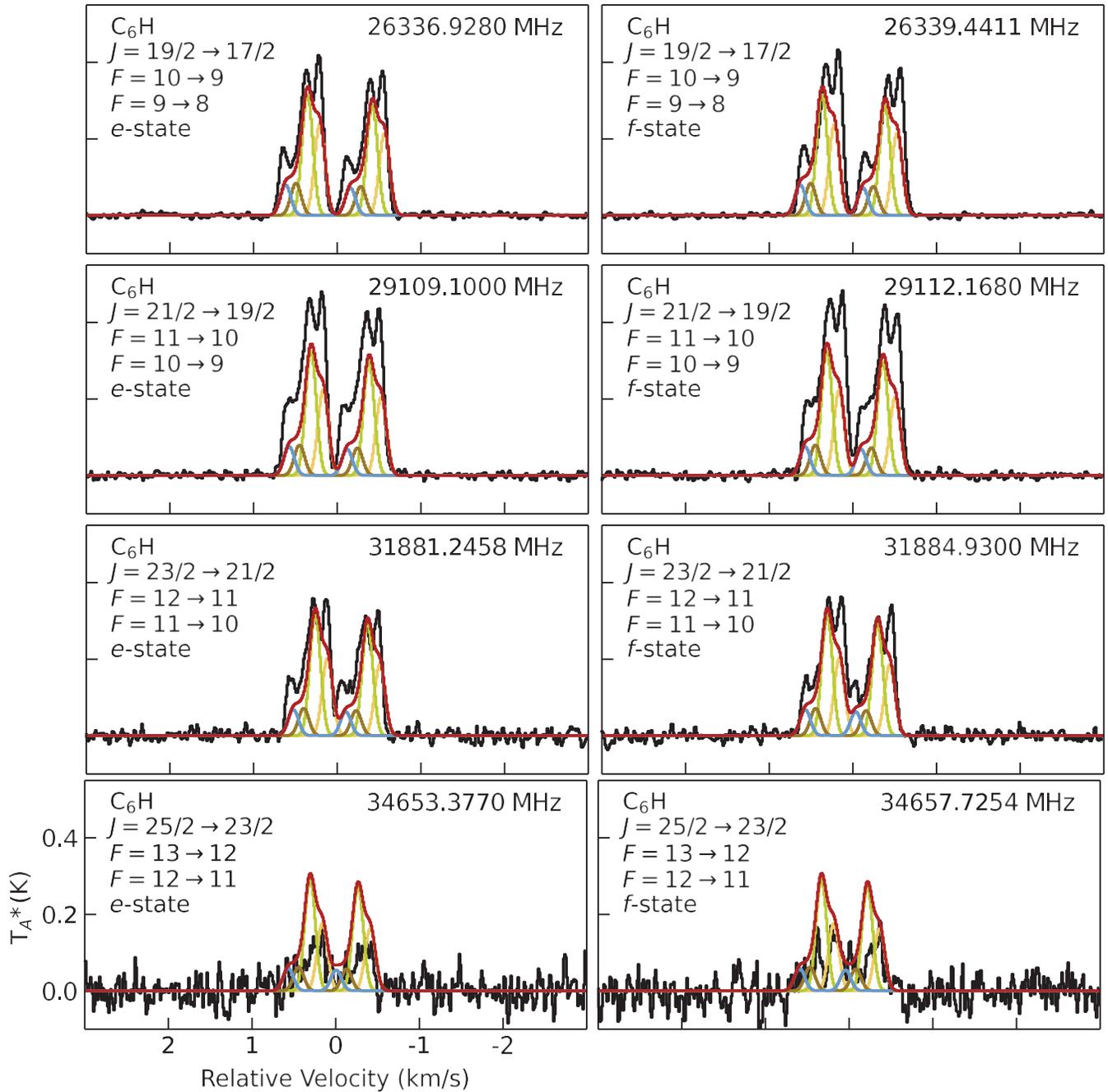

**Figure D3.** Individual lines of $C_6H$ detected in the GOTHAM observations (black). Simulations of $C_6H$ emission using the parameters given in Table D1 are shown in colors, with the total simulation in red. The quantum numbers for each transition are given in the upper left of each panel, and the central frequency of the window (in the sky frame) is given in the top right. Each window is 6.0 km s$^{-1}$ in total width.

### E. C₆H⁻ ANALYSIS

The best-fit parameters from the MCMC fit for C₆H⁻ are shown in Table E1. The corner plot from the analysis is shown in Fig. E1. Fig. E2 shows the individual lines of C₆H⁻ detected in the GOTHAM observations.

**Table E1.** $C_6H^-$ Values

| $v_{lsr}$ (km s$^{-1}$) | Size ($''$) | $N_T$ ($10^{10}$cm$^{-2}$) | $T_{ex}$ (K) | $\Delta V$ (km s$^{-1}$) |
|---|---|---|---|---|
| $5.646^{+0.009}_{-0.009}$ | $91^{+19}_{-13}$ | $8.76^{+1.49}_{-1.27}$ | | |
| $5.784^{+0.017}_{-0.018}$ | $32^{+4}_{-4}$ | $9.94^{+2.52}_{-2.24}$ | $4.32^{+0.22}_{-0.18}$ | $0.177^{+0.013}_{-0.012}$ |
| $5.905^{+0.024}_{-0.031}$ | $18^{+4}_{-4}$ | $6.66^{+4.95}_{-3.53}$ | | |
| $5.991^{+0.011}_{-0.010}$ | $636^{+249}_{-274}$ | $3.08^{+0.41}_{-0.43}$ | | |

$N_T$(Total): $2.84^{+0.58}_{-0.44} \times 10^{11}$ cm$^{-2}$

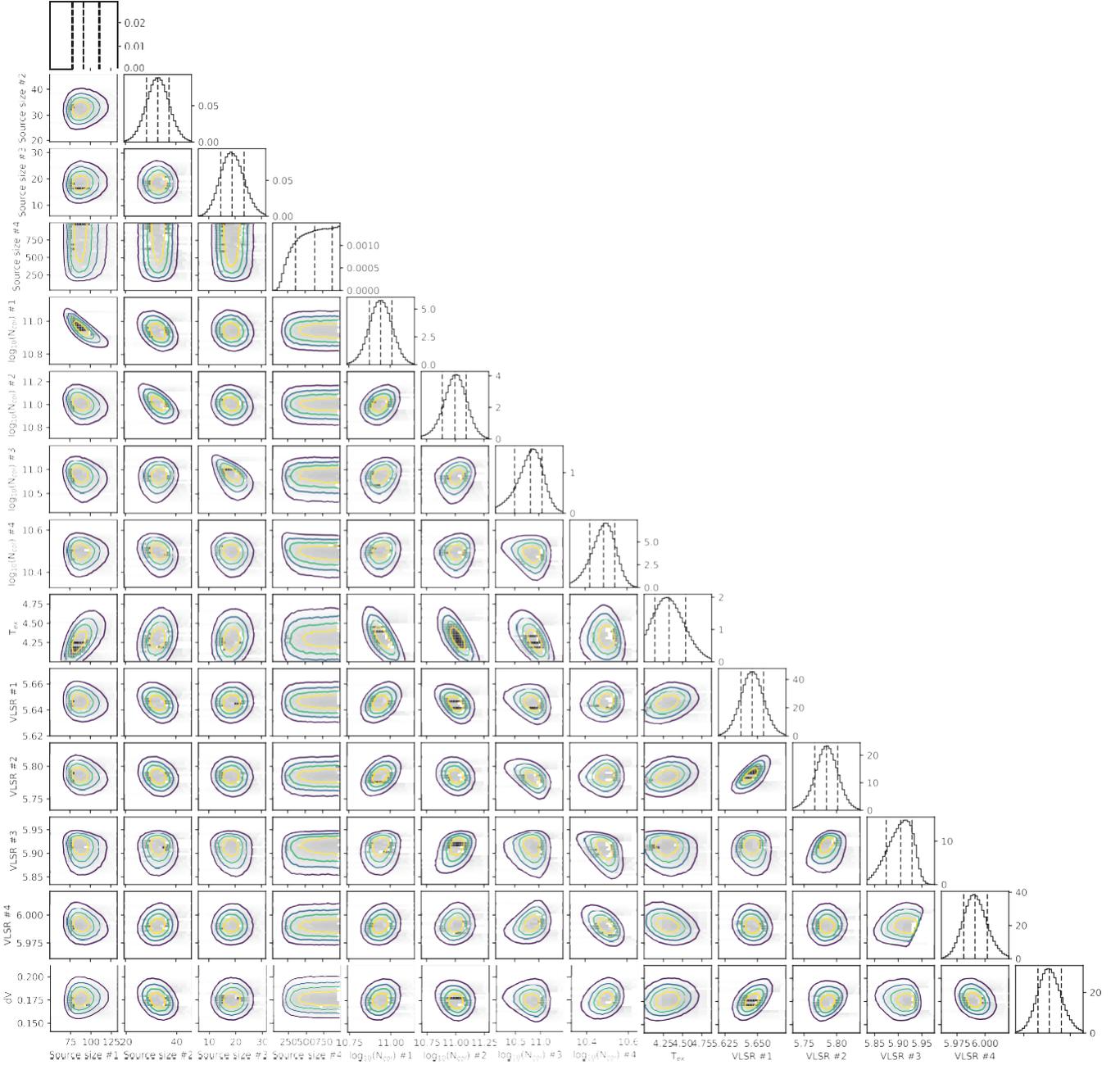

**Figure E1.** Corner plot for C$_6$H$^-$ showing parameter covariances and marginalized posterior distributions for the C$_6$H$^-$ MCMC fit. 16$^{th}$, 50$^{th}$, and 84$^{th}$ confidence intervals (corresponding to ±1 sigma for a Gaussian posterior distribution) are shown as vertical lines.

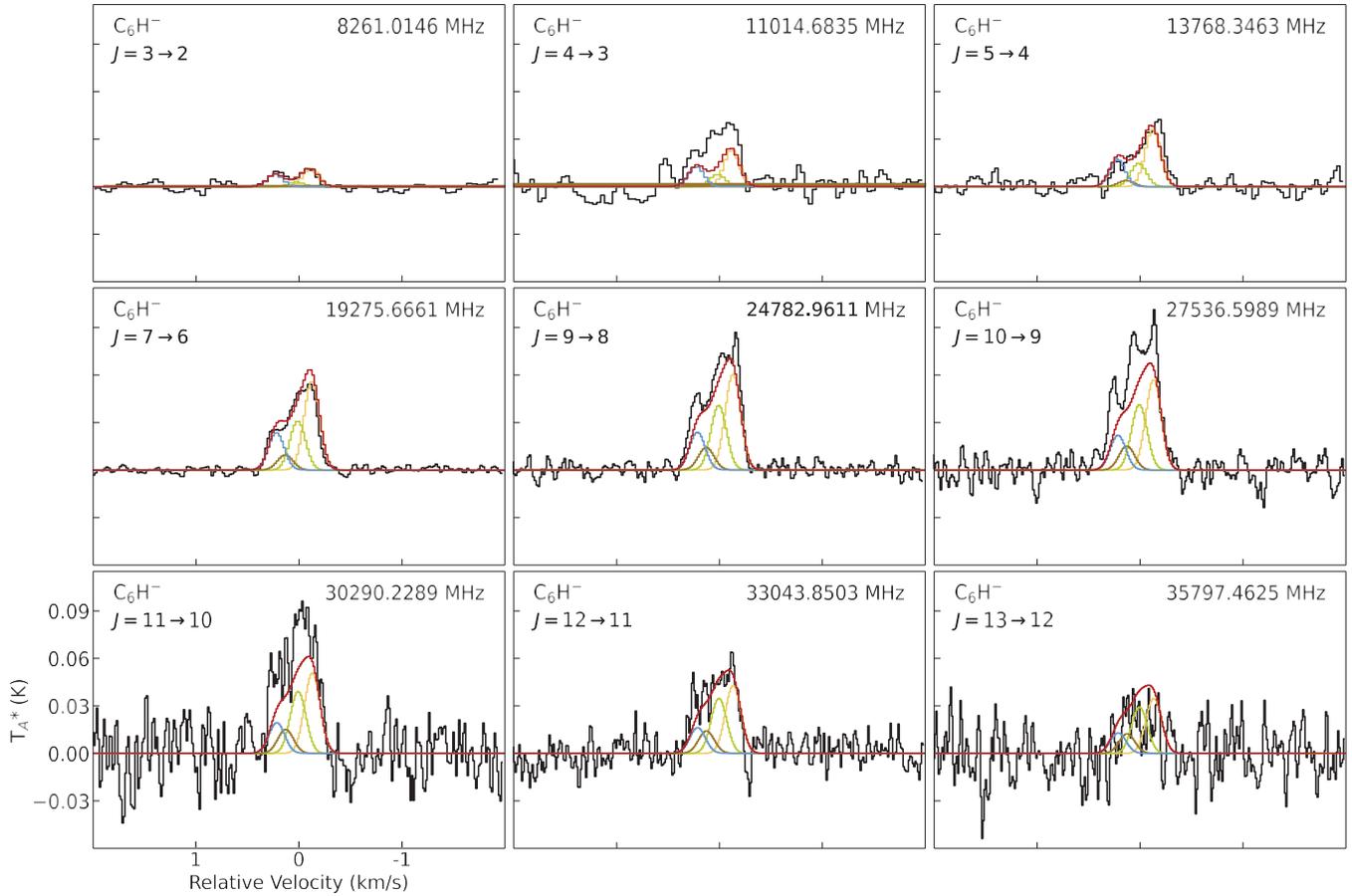

**Figure E2.** Individual lines of C₆H⁻ detected in the GOTHAM observations (black). Simulations of C₆H⁻ emission using the parameters given in Table E1 are shown in colors, with the total simulation in red. The quantum numbers for each transition are given in the upper left of each panel, and the central frequency of the window (in the sky frame) is given in the top right. Each window is 4.0 km s⁻¹ in total width.

## F. C$_8$H ANALYSIS

The best-fit parameters from the MCMC fit for C$_8$H are shown in Table F1. The corner plot from the analysis is shown in Fig. F1. Fig. F2 shows the individual lines of C$_8$H detected in the GOTHAM observations.

**Table F1.** C$_8$H Values

| $v_{lsr}$ (km s$^{-1}$) | Size ($''$) | $N_T$ ($10^{11}$cm$^{-2}$) | $T_{ex}$ (K) | $\Delta V$ (km s$^{-1}$) |
|---|---|---|---|---|
| $5.632^{+0.004}_{-0.004}$ | $696^{+209}_{-253}$ | $1.35^{+0.06}_{-0.07}$ | | |
| $5.775^{+0.007}_{-0.008}$ | $20^{+4}_{-3}$ | $5.00^{+2.05}_{-1.22}$ | $7.15^{+0.18}_{-0.18}$ | $0.159^{+0.010}_{-0.008}$ |
| $5.897^{+0.031}_{-0.043}$ | $654^{+237}_{-265}$ | $0.47^{+0.10}_{-0.14}$ | | |
| $6.010^{+0.027}_{-0.021}$ | $672^{+227}_{-267}$ | $0.44^{+0.15}_{-0.15}$ | | |

$N_T$(Total): $7.25^{+2.06}_{-1.23} \times 10^{11}$ cm$^{-2}$

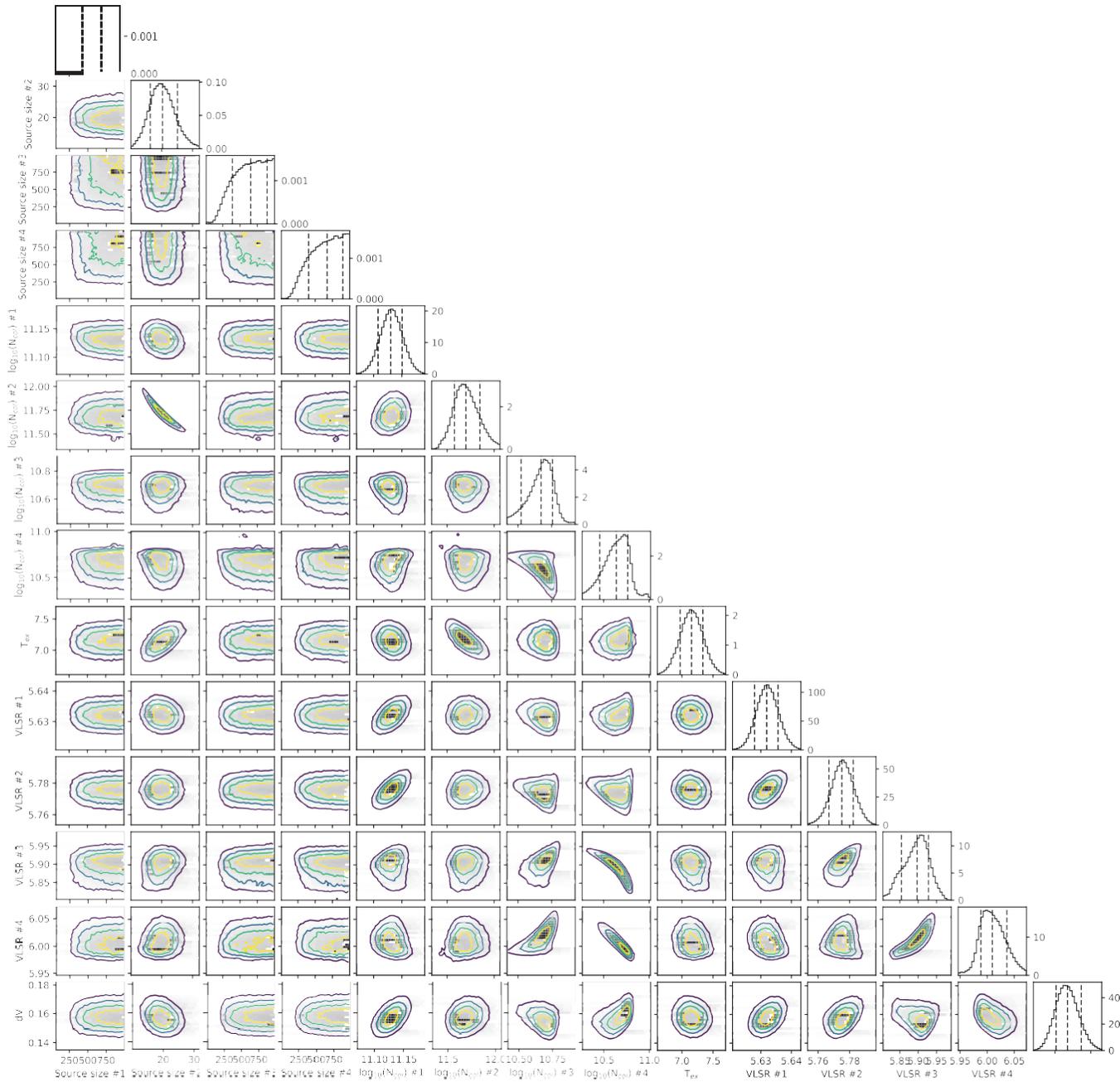

**Figure F1.** Corner plot for $C_8H$ showing parameter covariances and marginalized posterior distributions for the $C_8H$ MCMC fit. $16^{th}$, $50^{th}$, and $84^{th}$ confidence intervals (corresponding to $\pm 1$ sigma for a Gaussian posterior distribution) are shown as vertical lines.

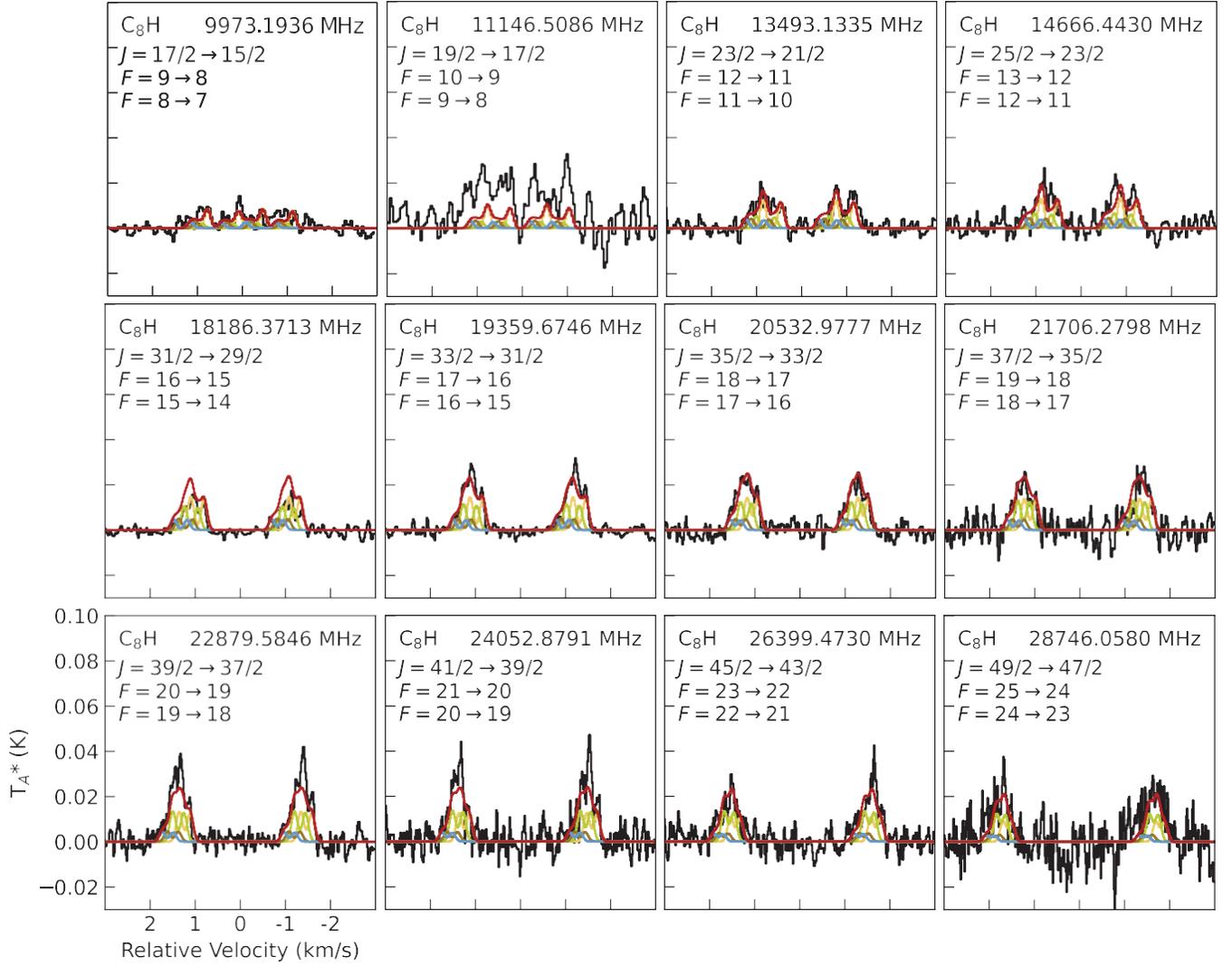

**Figure F2.** Individual lines of $C_8H$ detected in the GOTHAM observations (black). Simulations of $C_8H$ emission using the parameters given in Table F1 are shown in colors, with the total simulation in red. The quantum numbers for each transition are given in the upper left of each panel, and the central frequency of the window (in the sky frame) is given in the top right. Note that for $C_8H$, the two separated features seen in each window represent the *e* (lower frequency) and *f* (higher frequency) states; the two listed hyperfine components are blended into the features. Each window is $6.0\,km\,s^{-1}$ in total width.

## G. C$_8$H- ANALYSIS

The best-fit parameters from the MCMC fit for C$_8$H- are shown in Table G1. The corner plot from the analysis is shown in Fig. G1. Fig. G2 shows the individual lines of C$_8$H- detected in the GOTHAM observations.

**Table G1.** C$_8$H$^-$ Values

| $v_{lsr}$ (km s$^{-1}$) | Size ($''$) | $N_T$ ($10^{10}$ cm$^{-2}$) | $T_{ex}$ (K) | $\Delta V$ (km s$^{-1}$) |
|---|---|---|---|---|
| $5.646^{+0.004}_{0.004}$ | $87^{+51}_{-27}$ | $0.70^{+0.17}_{0.10}$ | | |
| $5.812^{+0.005}_{0.009}$ | $29^{+4}_{-4}$ | $2.20^{+0.74}$ | | |
| $5.896^{+0.037}_{0.030}$ | $18^{+5}_{-4}$ | $0.51^{+1.05}_{-0.38}$ | $6.17^{+0.30}_{-0.29}$ | $0.115^{+0.005}_{-0.005}$ |
| $6.021^{+0.005}_{0.005}$ | $12^{+7}_{-5}$ | $4.59^{+7.68}_{-2.50}$ | | |

$N_T$(Total): $8.00^{+7.79}_{-2.59} \times 10^{10}$ cm$^{-2}$

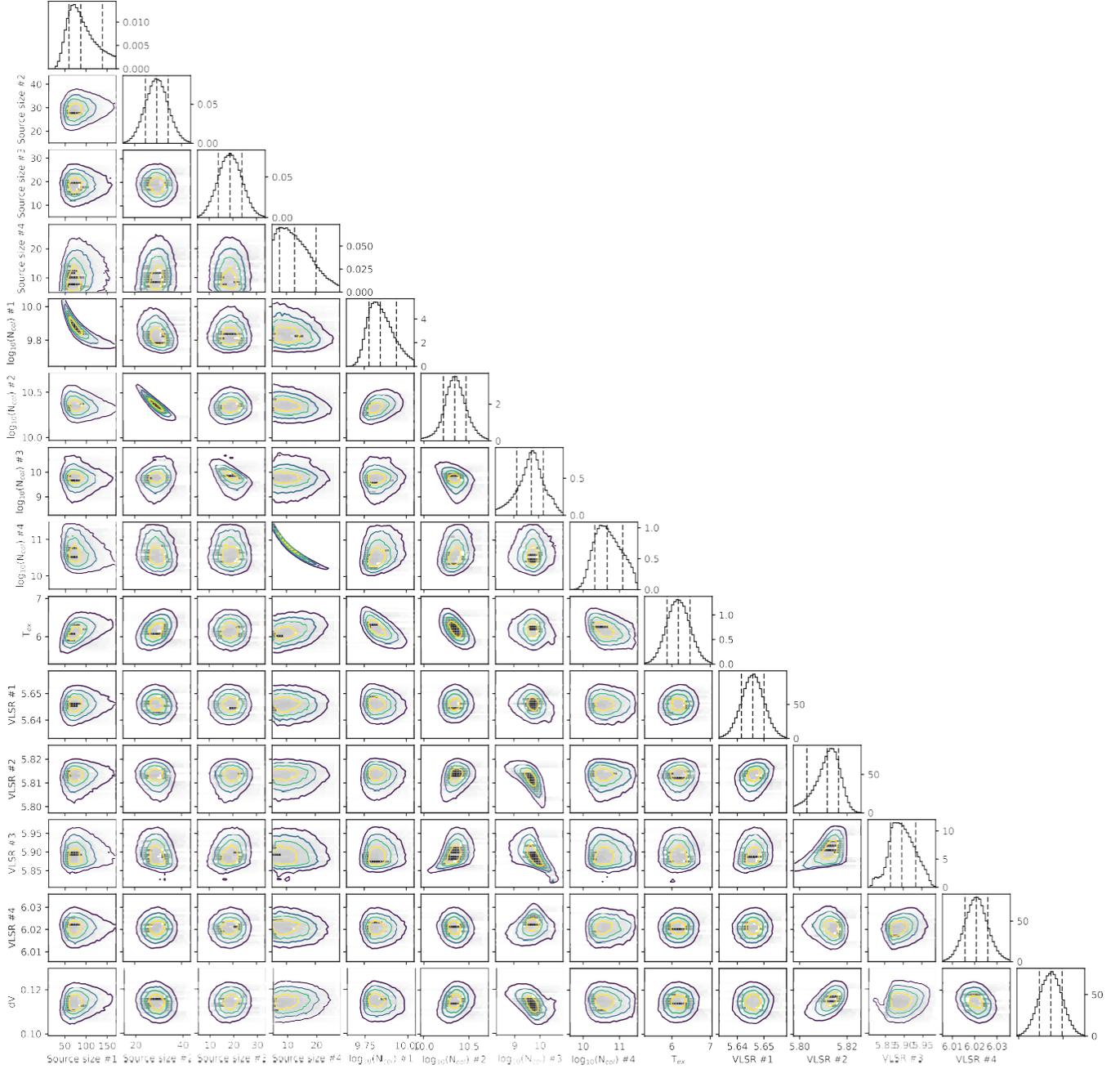

**Figure G1.** Corner plot for C$_8$H$^-$ showing parameter covariances and marginalized posterior distributions for the C$_8$H$^-$ MCMC fit. 16$^{th}$, 50$^{th}$, and 84$^{th}$ confidence intervals (corresponding to $\pm 1$ sigma for a Gaussian posterior distribution) are shown as vertical lines.

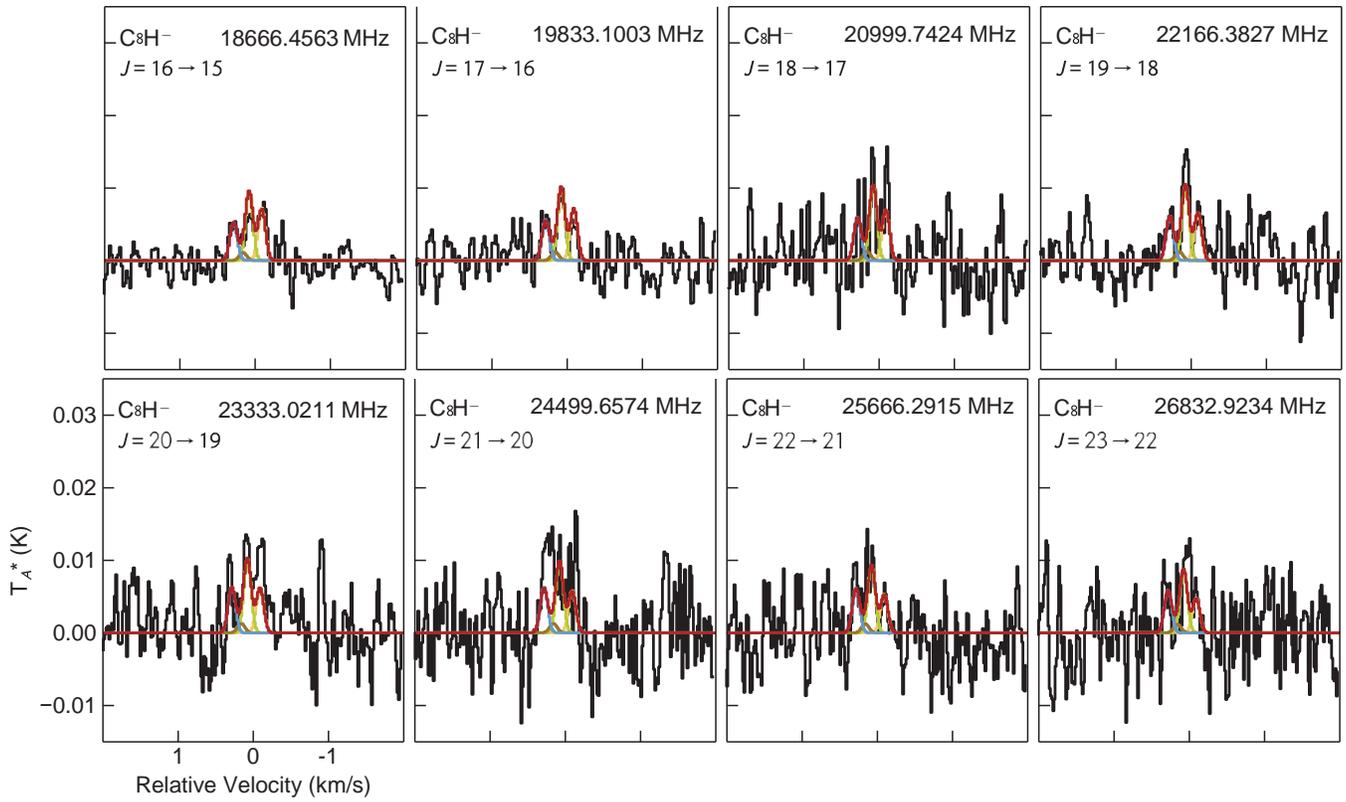

**Figure G2.** Individual lines of $C_8H^-$ detected in the GOTHAM observations (black). Simulations of $C_8H^-$ emission using the parameters given in Table G1 are shown in colors, with the total simulation in red. The quantum numbers for each transition are given in the upper left of each panel, and the central frequency of the window (in the sky frame) is given in the top right. Each window is 4.0 km s$^{-1}$ in total width.

## H. SPECTROSCOPY

The sources of the spectroscopic catalogs used for the analysis, as well as the literature references from which those catalogs were created, are provided in Table H1. The rotational partition function values for each of the molecules analyzed here are provided in Table H2 at the standard ste of temperatures from SPFIT/SPCAT.

**Table H1.** Sources for the spectroscopic catalogs used in the analysis for each molecule.

| Molecule | Catalog Source | Lab Ref. |
|---|---|---|
| $C_4H$ | This work | Gottlieb et al. 1983, this work |
| $C_4H^-$ | CDMS | Gupta et al. 2007; McCarthy & Thaddeus 2008; Amano 2008 |
| $C_6H$ | This work | Gottlieb et al. 2010 |
| $C_6H^-$ | CDMS | McCarthy et al. 2006 |
| $C_8H$ | CDMS | McCarthy et al. 1996, 1999 |
| $C_8H^-$ | CDMS | Gupta et al. 2007 |
| $C_{10}H$ | CDMS | Gottlieb et al. 1998b |
| $C_{10}H^-$ | This work | – |

**Table H2.** Values of the rotational partition function used in the analysis for each of the molecules at the standard set of temperatures from SPFIT/SPCAT.

| T (K) | $C_4H$ | $C_4H^-$ | $C_6H$ | $C_6H^-$ | $C_8H$ | $C_8H^-$ | $C_{10}H$ | $C_{10}H^-$ |
|---|---|---|---|---|---|---|---|---|
| 9.375 | 18.9084 | 42.3000 | 623.9222 | 142.2101 | 1407.8848 | 335.2048 | 1336.6177 | 651.7587 |
| 18.75 | 53.8950 | 84.2651 | 1478.7845 | 284.0880 | 3275.2896 | 670.0792 | 3029.9181 | 1303.1895 |
| 37.5 | 88.9165 | 168.1992 | 3503.2976 | 567.8499 | 7869.6366 | 1339.8382 | 7308.6095 | 2605.6555 |
| 75.0 | 123.9430 | 336.0788 | 7850.9514 | 1135.3965 | 18005.0283 | 2679.3957 | 17004.5164 | 5146.2785 |
| 150.0 | 165.5392 | 671.8809 | 16755.0177 | 2270.5807 | 38992.1292 | 5358.6689 | 37330.0583 | 9254.1018 |
| 225.0 | 211.5154 | 1007.7399 | 25714.4862 | 3405.8862 | 60175.2140 | 8038.1533 | 57920.6283 | 11998.1575 |
| 300.0 | 264.0606 | 1343.6560 | 34689.6933 | 4541.3128 | 81412.8992 | 10717.8486 | 78585.6294 | 13864.1185 |

## I. CHEMICAL MODELING TIME DEPENDENCE

Figure I1 shows the time-dependence of the simulated abundances within the nautilus chemical models in relation to the observed values, shown as hashed horizontal boxes. As it can be seen, the relative trend lines of these species can be quite time dependent, the time of peak abundance is strongly dependent on carbon chain length. For ease of comparison to previous studies of carbon-chains in TMC-1, we adopt the same source age as Siebert et al. (2022) of $2.5 \times 10^5$ years, as also shown in Figure I1.

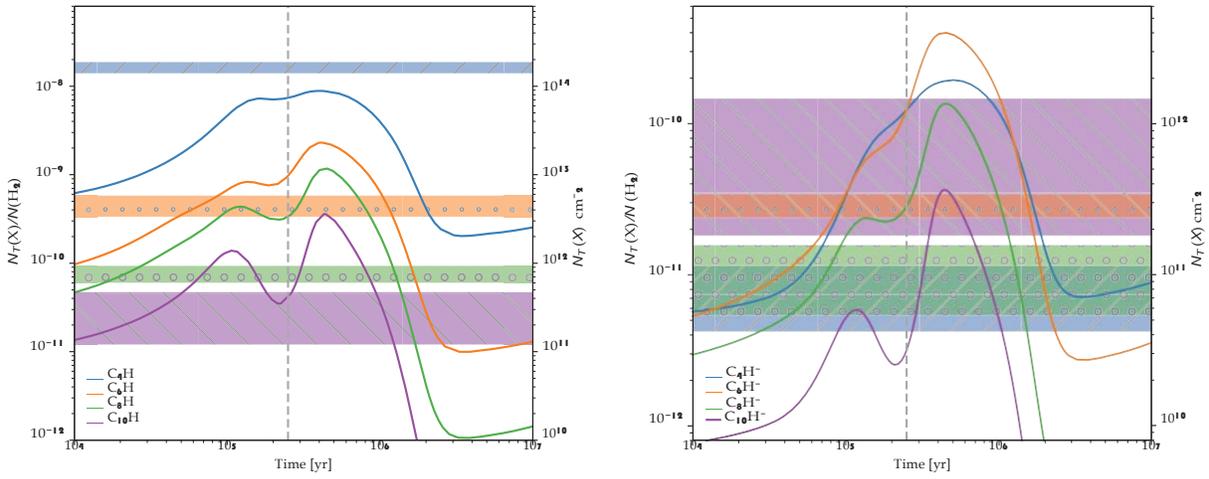

**Figure I1.** Simulated gas-phase abundance and column densities of the C$_n$N (*left*) and C$_n$H$^-$ (*right*) families from nautilus chemical models in comparison to the observed values with uncertainties as a horizontal bars with hashed patterns. The time used in Figure 4 is shown as a vertical dashed gray line. The same hash and color scheme is used for molecules containing the same number of carbon atoms